# A Shorter $^{146}$Sm Half-Life Measured and Implications for $^{146}$Sm-$^{142}$Nd Chronology in the Solar System

N. Kinoshita[1], M. Paul[2*], Y. Kashiv[3*], P. Collon[3], C. M. Deibel[4,5], B. DiGiovine[4], J. P. Greene[4], D. J. Henderson[4], C. L. Jiang[4], S. T. Marley[4], T. Nakanishi[6], R. C. Pardo[4], K. E. Rehm[4], D. Robertson[3], R. Scott[4], C. Schmitt[3], X. D. Tang[3], R. Vondrasek[4], A. Yokoyama[6]

[1]*Research Facility Center for Science and Technology, U. of Tsukuba, Japan*
[2]*Racah Institute of Physics, Hebrew University, Jerusalem, Israel 91904*
[3]*Department of Physics, University of Notre Dame, Notre Dame, IN 46556-5670*
[4]*Physics Division, Argonne National Laboratory, Argonne , IL 60439*
[5]*Joint Institute for Nuclear Astrophysics, Michigan State University, East Lansing, MI 46624*
[6]*Faculty of Chemistry, Institute of Science and Engineering, Kanazawa University, Japan*

**The extinct *p*-process nuclide $^{146}$Sm serves as an astrophysical and geochemical chronometer through measurements of isotopic anomalies of its α-decay daughter $^{142}$Nd. Based on analyses of $^{146}$Sm/$^{147}$Sm alpha-activity and atom ratios, we determined the half-life of $^{146}$Sm to be 68 ± 7 (1$\sigma$) million years (Ma), which is shorter than the currently used value of 103 ± 5 Ma. This half-life value implies a higher initial $^{146}$Sm abundance in the early solar system, ($^{146}$Sm/$^{144}$Sm)$_0$ = 0.0094±0.0005 (2$\sigma$), than previously estimated. Terrestrial, Lunar and Martian planetary silicate mantle differentiation events dated with $^{146}$Sm-$^{142}$Nd converge to a shorter time span and in general to earlier times, due to the combined effect of the new $^{146}$Sm half-life and ($^{146}$Sm/$^{144}$Sm)$_0$ values.**

______________________________________________

*To whom correspondence should be addressed. E-mail: paul@vms.huji.ac.il (M.P.); ykashiv@nd.edu (Y.K.)

The α decay of $^{146}$Sm serves as a clock for determining the chronology of solar system formation (*1*) and planetary differentiation (*2*) and can reveal insights into *p*-process nucleosynthesis of solar $^{146}$Sm (*3*). The correlation between anomalies in isotopic abundances of $^{142}$Nd and the Sm content, first observed (*4*) in meteorites, provided evidence for extinct $^{146}$Sm (entirely decayed to $^{142}$Nd). Neodymium-142 anomalies relative to the chondritic uniform reservoir (CHUR) observed in meteorite parent bodies (*5-7*), Earth (*8-11*), Moon (*12-14*) and Mars (*15-18*), primarily positive, are attributed to a small fractionation during partial melting or solidification favoring Sm in the solid phase in the silicate mantle owing to the slightly higher incompatibility of Nd, while $^{146}$Sm was still live (not entirely decayed). Caro *et al*. (*18*) inferred in addition that the Sm/Nd ratio of bulk Earth, Moon and Mars is 5-10 % higher than CHUR, which accounts for part of the positive anomalies, possibly due to fractionation and collisional erosion (*2, 14, 18, 19*).

The half-life of $^{146}$Sm, which sets the scale of the $^{146}$Sm-$^{142}$Nd clock, was previously measured four times (*20-23*) between 1953 and 1987 and its currently adopted value, derived from the works of Friedman *et al*. (*22*) and Meissner *et al*. (*23*), is 103 ± 5 Ma. Considering the range of measured $^{146}$Sm half-life ($t_{1/2}^{146}$) values (~ 50 Ma (*20*), 74±15 Ma (*21*), 103 ± 5 Ma (*22,23*)) and its importance in solar system chronology (*2*), we re-determined $t_{1/2}^{146}$ by measuring the α-activity ratio and atom ratio of $^{146}$Sm to naturally-occurring $^{147}$Sm in activated samples of $^{147}$Sm and using the equation $t_{1/2}^{146} = \frac{A_{147}}{A_{146}} \times \frac{N_{146}}{N_{147}} \times t_{1/2}^{147}$. Here, $t_{1/2}^{147}$ denotes the $^{147}$Sm alpha-decay half-life, 107.0 ± 0.9 billion years (Ga) (*24*), and $A_A$, $N_A$ the alpha activity and atom number of $^A$Sm in the sample, respectively. The ratio measurement eliminates most systematic uncertainties in determining the α activity due to detector efficiency and geometrical acceptance. We produced three independent $^{146}$Sm source materials by activating isotopically enriched $^{147}$Sm targets via the following nuclear reactions: (*i*) $^{147}$Sm($\gamma$,n)$^{146}$Sm (using 50-MeV electron bremsstrahlung radiation); (*ii*) 21-MeV proton irradiation through the $^{147}$Sm(p,2n$\varepsilon$)$^{146}$Sm reaction and (*iii*) fast-neutron activation $^{147}$Sm(n,2n)$^{146}$Sm (*25*). We prepared spectroscopic alpha sources (20-100 μg) from the three activations and counted them during several months using a silicon surface-barrier detector at Kanazawa University (Fig. 1, *25*), determining thus the α-activity ratio. For the determination of the $^{146}$Sm/$^{147}$Sm atom ratio, we used accelerator mass spectrometry (AMS) at the ATLAS facility (Argonne National Laboratory) because of the need to discriminate isobaric $^{146}$Nd interferences, observed to be critical in our experiment when using thermal-ionization mass spectrometry (*26*). The AMS isobaric separation was accomplished by the combined action of the high energy of the Sm ions after acceleration and use of a gas-filled magnetic spectrograph (*27, 28,* Fig.1). Two methods were used to determine the $^{146}$Sm/$^{147}$Sm atom ratio: $^{146}$Sm counting rate versus either $^{147}$Sm$^{22+}$ charge current or quantitatively attenuated $^{147}$Sm counting rate, giving consistent results (see (*25, 29*) for details on sample preparation and methods of measurement).

Our measured $^{146}$Sm half-life is shorter than the adopted value by a factor 0.66 ± 0.07 (1σ) (unweighted average and standard deviation) (Fig. 2). This corresponds to a $^{146}$Sm half-life of 68 ± 7 Ma (1σ); the systematic error (1 %) associated with the $^{147}$Sm half-life (*24*) is small compared to the random errors. Our value is shorter by ~30% than the values (102.6 ± 4.8 Ma) and (103.1 ± 4.5) Ma from (*22, 23*), respectively, and consistent (within larger uncertainties) with the two earlier works of Dunlavey and Seaborg (*20*) and Nurmia *et al*. (*21*), ~50 Ma and 74 ± 15 Ma, respectively. Different from these previous measurements, our AMS determination of $^{146}$Sm is free of isobaric contributions ($^{146}$Nd) which can cause contamination in rare-earth chemistry and mass spectrometry. The determination of an isotopic ratio ($^{146}$Sm/$^{147}$Sm) makes our experiment also insensitive to systematic effects inherent to measuring an absolute number of $^{146}$Sm atoms as was done in previous measurements.

Our $^{146}$Sm half-life value implies a reevaluation of the initial ($^{146}$Sm/$^{144}$Sm)$_0$ ratio in the solar system and reinterpretation of the $^{146}$Sm source and dating of planetary differentiation events (Table 1). Although evidence for extinct $^{146}$Sm was found in many meteorites, the majority show evidence of chemical and isotopic re-equilibration which potentially altered the initial ratio (*5-7*). In a recent study, Boyet *et al.* (*30*) selected a number of meteorites which appear to have remained closed Sm-Nd systems. Using the individual $^{146}$Sm/$^{144}$Sm ratios of the meteorites and $^{147}$Sm-$^{143}$Nd ages, they re-determined the initial solar system value ($^{146}$Sm/$^{144}$Sm)$_0$ = 0.0085±0.0007 (2σ) (compared to 0.008±0.001 (2σ) used in most studies (*6*)). As a basis for comparison, we fit the same set of data using our measured $^{146}$Sm half-life and derive an initial ratio ($^{146}$Sm/$^{144}$Sm)$_0$ = 0.0094±0.0005 (2σ) (Fig. 3). A fit taking into account uncertainties in both age and isotopic ratio leads, however, to the less precise value $0.0094^{+0.0018}_{-0.0015}$ (2σ). The determination of an experimental $^{146}$Sm decay curve from meteoritic data would require precision in $^{146}$Sm abundances beyond present capability or more data on meteorites dated to < 4,500 Ma before present. The initial ($^{146}$Sm/$^{144}$Sm)$_0$ ratio results from the decay of $^{146}$Sm in the interstellar medium (ISM) during an isolation interval Δ and both these quantities depend on the $^{146}$Sm half-life. The initial ratio can be expressed (*1, 31*) in terms of the ISM abundance ($^{146}$Sm/$^{144}$Sm)$_{ISM}$ as $(\frac{^{146}Sm}{^{144}Sm})_0 = (\frac{^{146}Sm}{^{144}Sm})_{ISM} \exp(-\Delta/\tau) = \frac{P^{146}}{P^{144}} \kappa \frac{\tau}{T} \exp(-\Delta/\tau)$ (1), where $P^{146}/P^{144}$ denotes the $^{146,144}$Sm *p*-process production ratio (taken as ~ 1), τ = 98 Ma is the $^{146}$Sm mean life determined in this work, T ~ 10 Ga the presolar age of the Galaxy and κ = p(T)/<p> the ratio of *p*-process rate just before solar system formation to the average *p*-process rate (κ ranging from 1 for the uniform-production (closed box) model (*1*) to ~ 2.7 for an open-box model (*31*) with Galactic-disk enrichment in low-metallicity gas). The isolation interval Δ calculated from eq. (1) with our half-life and ($^{146}$Sm/$^{144}$Sm)$_0$ values is reduced by factor of ~ 2.5 - 20 from previous estimates to ~5 Ma in the closed-box model and ~100 Ma in the open-box model.

The $^{146}$Sm decay curve using $t^{146}_{1/2}$ = 68 Ma, ($^{146}$Sm/$^{144}$Sm)$_0$ = 0.0094 and the two other decay curves with $t^{146}_{1/2}$ = 103 Ma (with ($^{146}$Sm/$^{147}$Sm)$_0$ = 0.0085 (*30*) and 0.008 (*6*)) intersect in the range of 30-50 Ma (Fig. 3). This reduces the age of late events (≥ 50 Ma) (age is measured in this work relative to the birth of the solar system) but has a minor effect on earlier events (Table 1). Samarium-146 observations in terrestrial samples can be divided into two groups: (i) most terrestrial rocks display a $^{142}$Nd/$^{144}$Nd ratio higher by ~18 parts per million (ppm) than CHUR (*9*) and (ii) anomalies in the $^{142}$Nd/$^{144}$Nd ratio relative to the terrestrial standard, both positive, in rocks from Greenland and Australia (*2,10*), and negative, in rocks from Northern Quebec (*2,11*). Two explanations were offered for the (*i*) anomalies, that the mantle had to differentiate within ≤ 30 Ma (*2, 9*) or that Earth (and the Moon and Mars) has a bulk Sm/Nd ratio 5-10% higher than CHUR (*2, 14, 18*) and neither is affected by the new values presented here. At the same time, ages derived from (*ii*) change significantly. One such example is the differentiation age of the depleted mantle source of Archean rocks from Isua (Greenland) (*2*): the interpretation of this array as an isochron (*2*) leads now to an age of 120 Ma rather than 170 Ma. The revised age improves the agreement with the 70-170 Ma $^{176}$Lu-$^{176}$Hf differentiation age derived from the Hadean Jack Hills zircons (Australia) (*2*), strongly suggesting that one global differentiation event is sampled in both cases. Another example is the isochron which dates the differentiation of the enriched mantle source of rocks from Quebec (*11*), whose estimated age decreases from $287^{+81}_{-53}$ to $205^{+54}_{-35}$ Ma. An array of Lunar rocks (*12, 13*) has been similarly interpreted as an isochron (*2*) which dates the solidification of the Lunar magma ocean. The isochron-inferred age decreases with the new values from 242 ± 22 to 170 ± 15 Ma. A recent study by Borg *et al.* (*32*) dated a Lunar sample, ferroan anorthosite (FAN) 60025, by three methods, Pb-Pb, $^{147}$Sm-$^{143}$Nd and $^{146}$Sm-$^{142}$Nd. The Pb-Pb and $^{147}$Sm-$^{143}$Nd systems gave consistent ages of 208.8 ± 2.4 and 201 ± 11 Ma, respectively, but the $^{146}$Sm-$^{142}$Nd age, $250^{+38}_{-30}$ Ma, showed a discrepancy. Our value of $^{146}$Sm half-life revises the latter to $175^{+25}_{-20}$ Ma (Table 1) and removes most of this discrepancy, bringing the $^{146}$Sm-$^{142}$Nd age into agreement with the $^{147}$Sm-$^{143}$Nd age. This is an important point considering that FAN 60025 is now the only rock dated precisely enough with the different systems to trace their decay. The new $^{146}$Sm age of FAN60025 remains in accord with the revised younger age of the Lunar array isochron.

**References and Notes**

1. G. J. Wasserburg, M. Busso, R. Gallino, K. M. Nollett, Short-lived nuclei in the early Solar System: Possible AGB sources. *Nucl. Phys. A* **777**, 5–69 (2006).

2. G. Caro, Early silicate earth differentiation. *Annu. Rev. Earth Planet. Sci.*, **39**, 31-58 (2011).

3. J. Audouze, D. N. Schramm, $^{146}$Sm: A chronometer for *p*-process nucleosynthesis. *Nature* **237**, 447-449 (1972).

4. G. W. Lugmair, K. Marti, Sm-Nd-Pu time pieces in the Angra dos Reis meteorite. *Earth Planet. Sci. Lett.* **35**, 273-284 (1977).

5. A. Prinzhofer, D. A. Papanastassiou, G. J. Wasserburg, Samarium-neodymium evolution of meteorites. *Geochim. Cosmochim. Acta* **56**, 797-815 (1992).

6. L. E. Nyquist, B. Bansal, H. Wiesmann, C.-Y. Shih, Neodymium, strontium and chromium isotopic studies of the LEW86010 and Angra dos Reis meteorites and the chronology of the angrite parent body. *Meteoritics* **29**, 872–885 (1994).

7. B. W. Stewart, D. A. Papanastassiou, G. J. Wasserburg, Sm–Nd chronology and petrogenesis of mesosiderites. *Geochim. Cosmochim. Acta* **58**, 3487–3509 (1994).

8. C. L. Harper, Jr., S. B. Jacobsen, Evidence from coupled $^{147}$Sm–$^{143}$Nd and $^{146}$Sm–$^{142}$Nd systematics for very early (4.5 Gyr) differentiation of the Earth's mantle. *Nature* **360**, 24-31 (1992).

9. M. Boyet, R. W. Carlson, $^{142}$Nd evidence for early (> 4.53 Ga) global differentiation of the silicate Earth. *Science* **309**, 576–581 (2005).

10. V. C. Bennett, A. D. Brandon, A. P. Nutman, Coupled $^{142}$Nd-$^{143}$Nd isotopic evidence for Hadean mantle dynamics. *Science* **318**, 1907–1910 (2007).

11. J. O'Neil, R. W. Carlson, D. Francis, R. K. Stevenson, Neodymium-142 evidence for Hadean mafic crust. *Science* **321**, 1828–1831 (2008).

12. L. E. Nyquist *et al.*, $^{146}$Sm–$^{142}$Nd formation interval for the lunar mantle. *Geochim. Cosmochim. Acta* **59**, 2817–2837 (1995).

13. M. Boyet, R. W. Carlson, A highly depleted moon or a non-magma ocean origin for the lunar crust? *Earth Planet. Sci. Lett.* **262**, 505–516 (2007).

14. A. D. Brandon, T. J. Lapen, V. Debaille, B. L. Beard, K. Rankenburg, C. Neal, Re-evaluating $^{142}$Nd/$^{144}$Nd in Lunar mare basalts with implications for the early evolution and bulk Sm/Nd of the Moon. *Geochim. Cosmochim. Acta* **73**, 6421–6445 (2009).

15. C. L. Harper, Jr., L. E. Nyquist, B. Bansal, H. Wiesmann, H., C.-Y. Shih, Rapid accretion and early differentiation of Mars indicated by $^{142}$Nd/$^{144}$Nd in SNC meteorites. *Science* **267**, 213–217 (1995).

16. C. N. Foley, M. Wadhwa, L. E. Borg, P. E. Janney, R. Hines, T. L. Grove, The early differentiation history of Mars from $^{182}$W-$^{142}$Nd isotope systematics in the SNC meteorites. *Geochim. Cosmochim. Acta* **69**, 4557–4571 (2005).

17. V. Debaille, A. D. Brandon, Q. Z. Yin, B. Jacobsen, Coupled $^{142}$Nd–$^{143}$Nd evidence for a protracted magma ocean in Mars. *Nature* **450**, 525–528 (2007).

18. G. Caro, B. Bourdon, A. N. Halliday, G. Quitté, Super-chondritic Sm/Nd ratios in Mars, the Earth and the Moon. *Nature* **452**, 336–339 (2008).

19. H. S. C. O'Neill, H. Palme, Collisional erosion and the nonchondritic composition of the terrestrial planets. *Philos. Trans. R. Soc. Ser. A* **366**, 4205–4238 (2008).

20. D. C. Dunlavey, G. T. Seaborg, Alpha activity of $^{146}$Sm as Detected with Nuclear Emulsions. *Phys. Rev.* **92** 206 (1953).

21. M. Nurmia, G. Graeffe, K. Valli, J. Aaltonen, Alpha activity of Sm-146. *Ann. Acad. Scient. Fenn.* A.VI.**148**, 1-8 (1964).

22. A. M. Friedman *et al.*, Alpha decay half-lives of $^{148}$Gd, $^{150}$Gd and $^{146}$Sm. *Radiochim. Acta* **5**, 192-194 (1966).

23. F. Meissner, W.-D. Schmidt-Ott, L. Ziegeler, Half-life and α-energy of $^{146}$Sm. *Z. Phys. A* **327**, 171-174 (1987).

24. K. Kossert, G. Joerg, O. Naehle, C. L. v. Gostomski, High-precision measurement of the half-life of $^{147}$Sm. *Appl. Rad. Isot.* **67**, 1702-1706 (2009).

25. Materials and methods are available on *Science* online

26. H. Tazoe, H. Obata, H. Amakawa, Y. Nozaki, T. Gamo, Precise determination of the cerium isotopic compositions of surface seawater in the Northwest Pacific Ocean and Tokyo Bay. *Mar. Chem.* **103**, 1-14 (2007).

27. M. Paul *et al.*, Heavy ion separation with a gas-filled magnetic spectrograph. *Nucl. Inst. Methods Phys. Res. A* **277**, 418-420 (1989).

28. N. Kinoshita *et al.*, Ultra-sensitive detection of *p*-process nuclide $^{146}$Sm produced by (γ,n), (p,pnε) and (n,2n) reactions. *J. Phys. G: Nucl. Part. Phys.* **35**, 014033 (2008).

29. N. Kinoshita et al., New AMS Method to Measure the Atom Ratio $^{146}$Sm/$^{147}$Sm for a Half-life Determination of $^{146}$Sm, *Nucl. Instr. Methods B* (2012), http://dx.doi.org/10.1016/j.nimb.2012.01.013.

30. M. Boyet, R. W. Carlson, M. Horan, Old Sm–Nd ages for cumulate eucrites and redetermination of the solar system initial $^{146}$Sm/$^{144}$Sm ratio. *Earth Planet. Sci. Lett*. **291**, 172–181 (2010).

31. N. Dauphas, T. Rauscher, B. Marty and L. Reisberg, Short-lived *p*-nuclides in the early solar system and implications on the nucleosynthetic role of X-ray binaries. *Nucl. Phys. A* **719**, 287c-295c (2003).

32. L. E. Borg, J. N. Connelly, M. Boyet, R. W. Carlson, Chronological evidence that the Moon is either young or did not have a global magma ocean. *Nature* **477**, 70-72 (2011).

33. N. Dauphas, A. Pourmand, Hf–W–Th evidence for rapid growth of Mars and its status as a planetary embryo. *Nature* **473**, 489–493 (2011).

34. N. Kinoshita *et al.*, Technological development for half-life measurement of $^{146}$Sm nuclide. *J. Nucl. Radiochem. Sci.* **8**, 109-112 (2007).

35. M. Schlapp *et al*., A new 14 GHz electron-cyclotron-resonance ion source for the heavy ion accelerator facility ATLAS. *Rev. Sci. Inst.* **69**, 631-633 (1998).

36. M. Paul *et al*., AMS of heavy elements with an ECR ion source and the ATLAS linear accelerator. *Nucl. Inst. Methods Phys. Res. B* **172**, 688-692 (2000).

37. J. K. Boehlke *et al*., Isotopic Compositions of the Elements, 2001. *J. Phys. Chem. Ref. Data*, **34**, 57-67 (2005).

38. T. Kleine, K. Mezger, C. Münker, H. Palme, A. Bischoff, $^{182}$Hf-$^{182}$W isotope systematics of chondrites, eucrites and Martian meteorites: Chronology of core formation and early mantle differentiation in Vesta and Mars. *Geochim. Cosmochim. Acta* **68**, 2935-2946 (2004).

39. G. Beard, M. L. Wiedenbeck, Natural Radioactivity of Sm$^{147}$. *Phys. Rev.* **95**, 1245-1246 (1954).

40. G. Beard, W. H. Kelly, The use of a samarium loaded liquid scintillator for the determination of the half-life of Sm$^{147}$. *Nucl. Phys.* **8**, 207-209 (1958).

41. R. D. MacFarlane, T. P. Kohman, Natural Alpha Radioactivity in Medium-Heavy Elements. *Phys. Rev.* **121**, 1758-1769 (1961).

42. P. M. Wright, E. P. Steinberg, L. E. Glendenin, Half-Life of Samarium-147. *Phys. Rev.* **123**, 205-208 (1961).

43. D. Donhoffer, Bestimmung der halbwertszeiten der in der natur vorkommenden radioaktiven nuklide Sm$^{147}$ und Lu$^{176}$ mittels flüssiger szintillatoren. *Nucl. Phys.* **50**, 489-496 (1964).

44. M. C. Gupta and R. D. MacFarlane, The natural alpha radioactivity of samarium. *J. Inorg. Nucl. Chem.* **32**, 3425-3432 (1970).

45. N. Kinoshita, A. Yokoyama, T. Nakanishi, Half-Life of Samarium-147. *J. Nucl. Radiochem. Sci*. **4**, 5-7 (2003)

46. J. Su *et al*.., Alpha decay half-life of $^{147}$Sm in metal samarium and Sm$_2$O$_3$. *Eur. Phys. J. A* **46**, 69–72 (2010).

**Acknowledgements**: We thank R. Carlson, W. Kutschera, K. Nollett and J. Schiffer for helpful discussions, and three anonymous reviewers for comments which greatly benefited the paper. This work is supported in part by Grant-in-Aid for Scientific Research Program of Japan Society for the Promotion of Science (20740161). This work is supported by the U.S. Department of Energy, Office of Nuclear Physics, under contract Nr. DE-AC02-06CH11357 and by the NSF JINA Grant Nr. PHY0822648.

**Supporting Online Material**: Materials and Methods

SOM text

Figs. S1 to S6

Table S1, S2

**Table 1: The effect of the shorter $^{146}$Sm half-life on estimated time of differentiation events.**

| Planetary body | Sample / Mantle differentiation event | Reference | Time after start of SS formation (Ma) | |
|---|---|---|---|---|
| | | | From ref. in col. 3[a] | Revised in present work[b] |
| Earth | Terrestrial rocks 18 ppm higher than CHUR / depleted–enriched reservoirs | 9 | $\leq 30$ | No change |
| Earth | Archean array, Isua, Greenland / depleted source | 2 | 170[c] | 120[d] |
| Earth | Nuvvuagittuq greenstone belt, Northern Quebec, Canada / enriched source[e] | 11 | $287^{+81}_{-53}$ | $205^{+54}_{-35}$ [f] |
| Moon | Lunar array / LMO solidification[g] | 2 | $242 \pm 22$ | $170 \pm 15$ [h] |
| Moon | FAN 60025 / LMO solidification[g,i] | 32 | $250^{+38}_{-30}$ | $175^{+25}_{-20}$ [j] |
| Mars | Nakhlites / solidification of depleted source | 16 | $8 - 25$ [k] | Favors young age[l] |
| Mars | Enriched shergottites / solidification of source | 17 | $\sim 110$ [m] | $\sim 90$ [n] |

[a] Derived in the original studies (col. 3) using $t^{146}_{1/2} = 103$ Ma (*22,23*), ($^{146}$Sm/$^{147}$Sm)$_0$ = 0.008 (*6*), $t_0 = 4{,}567$ Ma (see text).

[b] Derived using the value presented in this work.

[c] Using ($^{146}$Sm/$^{147}$Sm)$_0$ = 0.0085.

[d] Reinterpretation of Fig. 3 of Caro (*2*).

[e] Measured in faux-amphibolite and gabbro (> 3,800 Ma before present).

[f] Reinterpretation of Fig. 3 of O'Neil *et al.* (*11*).

[g] LMO = Lunar magma ocean.

[h] Reinterpretation of Fig. 8 of Caro (*2*).

[i] FAN = ferroan anorthosite.

[j] Reinterpretation of Fig. 2b of Borg *et al*. (*32*).

[k] Calculated a depleted source of $^{147}$Sm/$^{144}$Nd ~0.255-0.266 ($^{180}$Hf/$^{183}$W = ~22-43) for the Martian mantle source.

[l] Calculated a less depleted source of $^{147}$Sm/$^{144}$Nd ~0.245 ($^{180}$Hf/$^{183}$W ~ 20) for the Martian mantle source (Fig. S5, *25*). The young age may be in line with the recent finding that Mars accreted within ~ 4 Ma (*33*).

[m] Calculated assuming the data form a mixing line and the source has CHUR Sm/Nd ratio. See Fig. S5 (*25*). Alternative interpretation of the data line as an isochron and a source with super-chondritic Sm/Nd ratio gives an age of ~40 Ma (*18*).

[n] See Fig. S6 (*25*).

**Figure legends**

**Figure 1:** (left) Alpha spectra from $^{147}$Sm activated via $(\gamma,n)$, $(n,2n)$ and $(p,2n\varepsilon)$ reactions, determining the $^{146}$Sm/$^{147}$Sm activity ratio; (right) identification spectra measured for a $(n,2n)$ activated sample (upper right) and non-activated $^{nat}$Sm (lower right). The spectra display energy loss vs. position signals from a detector along the focal plane of a gas-filled magnetic spectrograph. The device separates in position (magnetic rigidity) $^{146}$Sm ions from $^{146}$Nd isobaric ions (originating from chemical impurities in the sample or in ion source structural material), owing to their different mean ionic charge state in the gas (*25*). The observed $^{126}$Xe$^{19+}$ group is transmitted by the accelerator due to its charge-to-mass ratio being quasi-degenerate with that of $^{146}$Sm$^{22+}$ and originates from residual gas in the ion source. In order to determine the $^{146}$Sm/$^{147}$Sm ratios (*25*), the count rate in the $^{146}$Sm group was normalized to $^{147}$Sm, alternately transported through the accelerator (see Fig. 2). The blank spectrum shown for the $^{nat}$Sm sample (lower right) corresponds to a ratio $^{146}$Sm/$^{147}$Sm $< 10^{-11}$.

**Figure 2:** Double ratios of the $^{146}$Sm/$^{147}$Sm atom ratio measured by AMS to that expected in the same sample from its $^{146}$Sm/$^{147}$Sm α activity ratio using the currently adopted $^{146}$Sm half-life, 103 Ma (*22, 23*). The double ratio is equivalent to the ratio of our measured $^{146}$Sm half-life to the currently adopted value (right vertical axis). The single dashed line (red) corresponds to a ratio of 1. The notations G-x, N-x, P-x on the horizontal axis denote various independently prepared samples from the bremsstrahlung, neutron and proton irradiations, respectively. The plotted value for each sample is the unweighted mean and standard deviation of the double ratios measured in repeat measurements of that sample (see Table S1 and Figs. S4, S5 for individual

values). The error bars represent random statistical errors in AMS ion counting and α activity, uncertainty in dilution factors and random errors in ion transmission. The unweighted mean and standard deviation of the double ratios shown, 0.66 ± 0.07, correspond to a $^{146}$Sm half-life of 68 ± 7 Ma. Diamond and circles represent values measured by normalizing $^{146}$Sm counts to $^{147}$Sm$^{22+}$ charge and squares to counts of a quantitatively attenuated $^{147}$Sm beam measured in the same detector as $^{146}$Sm ions, avoiding possible systematic uncertainties in the charge current measurement or in transmission efficiency between the Faraday cup and detector (*25, 29*).

**Figure 3:** Reinterpretation of the ($^{146}$Sm/$^{144}$Sm)$_0$ initial ratio measured in selected meteorites ((*30*) and references therein) plotted against the meteorite age after solar system formation in Ma (upper horizontal axis) or time in Ma before present (bp) (lower horizontal axis). The time of solar system formation is taken as 4568 Ma bp. The set of meteorites selected by Boyet *et al.* (*30*) as closed Sm-Nd systems is used here: 4 eucrites (Moore County, EET87520, Caldera, Binda), one mesosiderite (Mt. Padburry) and one angrite (LEW 86010). For the latter, a weighted average of the age and isotopic ratio of two measurements (*30*) was used here. Two additional eucrites (Y980318/433), with lower isotopic ratios, are shown but not included in the calculation (*30*). The data are fitted by a new decay curve (solid blue line) using the $^{146}$Sm half-life value measured in this work ($t^{146}_{1/2}$ = 68 Ma) and yield an initial solar system ratio $r_0$= ($^{146}$Sm/$^{144}$Sm)$_0$ = 0.0094 ± 0.0005 (2σ), compared to the curve from Boyet *et al.* (*30*) (dashed) with $t^{146}_{1/2}$ = 103 Ma, $r_0$= 0.0085 ± 0.0007 (2σ) and that used in most studies (*6*) (dotted) with $t^{146}_{1/2}$ = 103 Ma, $r_0$= 0.008 ± 0.001. Both the $^{146}$Sm half-life value of 68 Ma and the initial solar system ratio $r_0$= ($^{146}$Sm/$^{144}$Sm)$_0$ = 0.0094 determined in this work are used for the age determination of differentiation processes (Table 1).

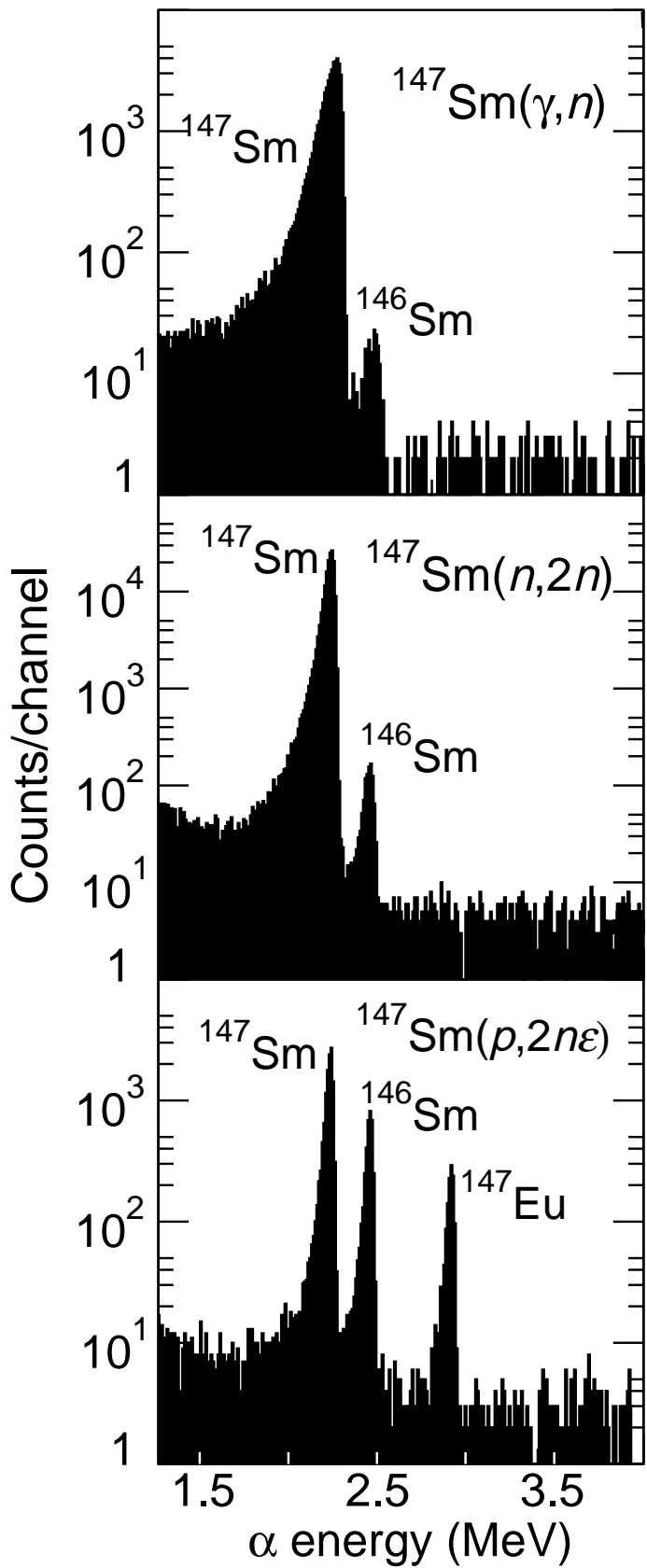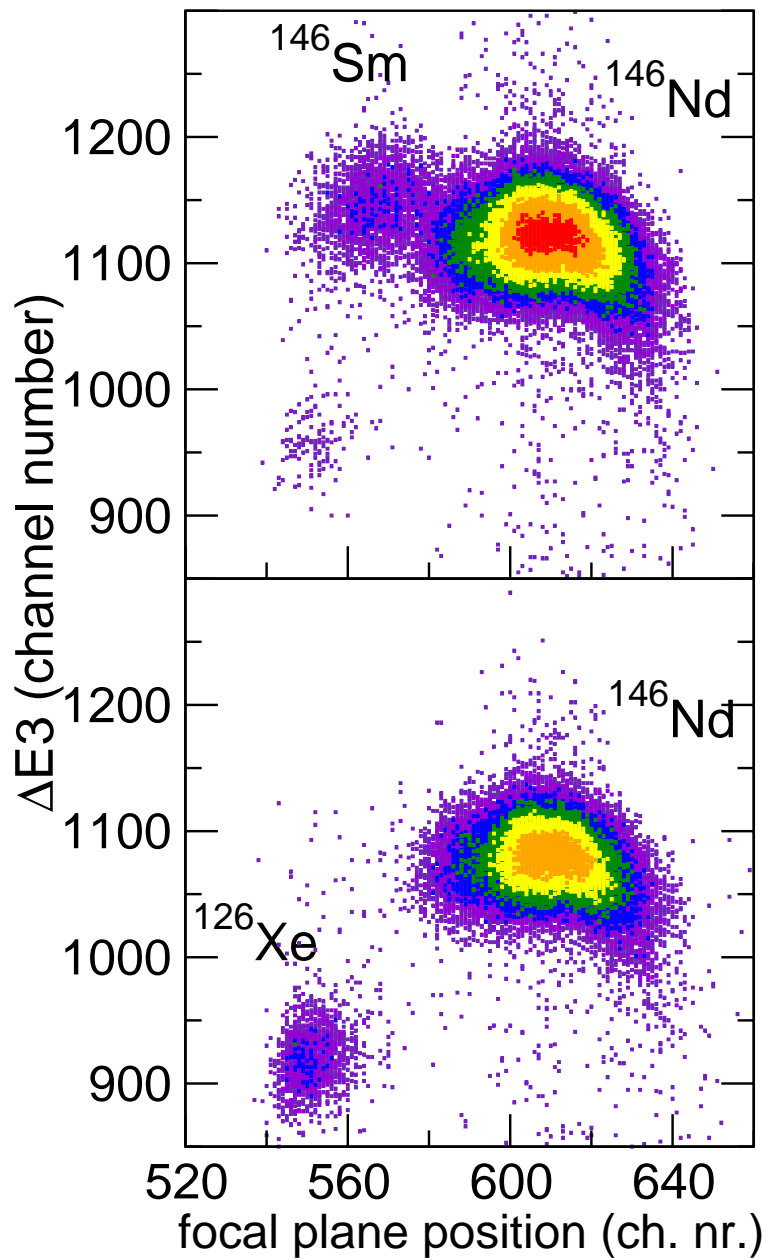

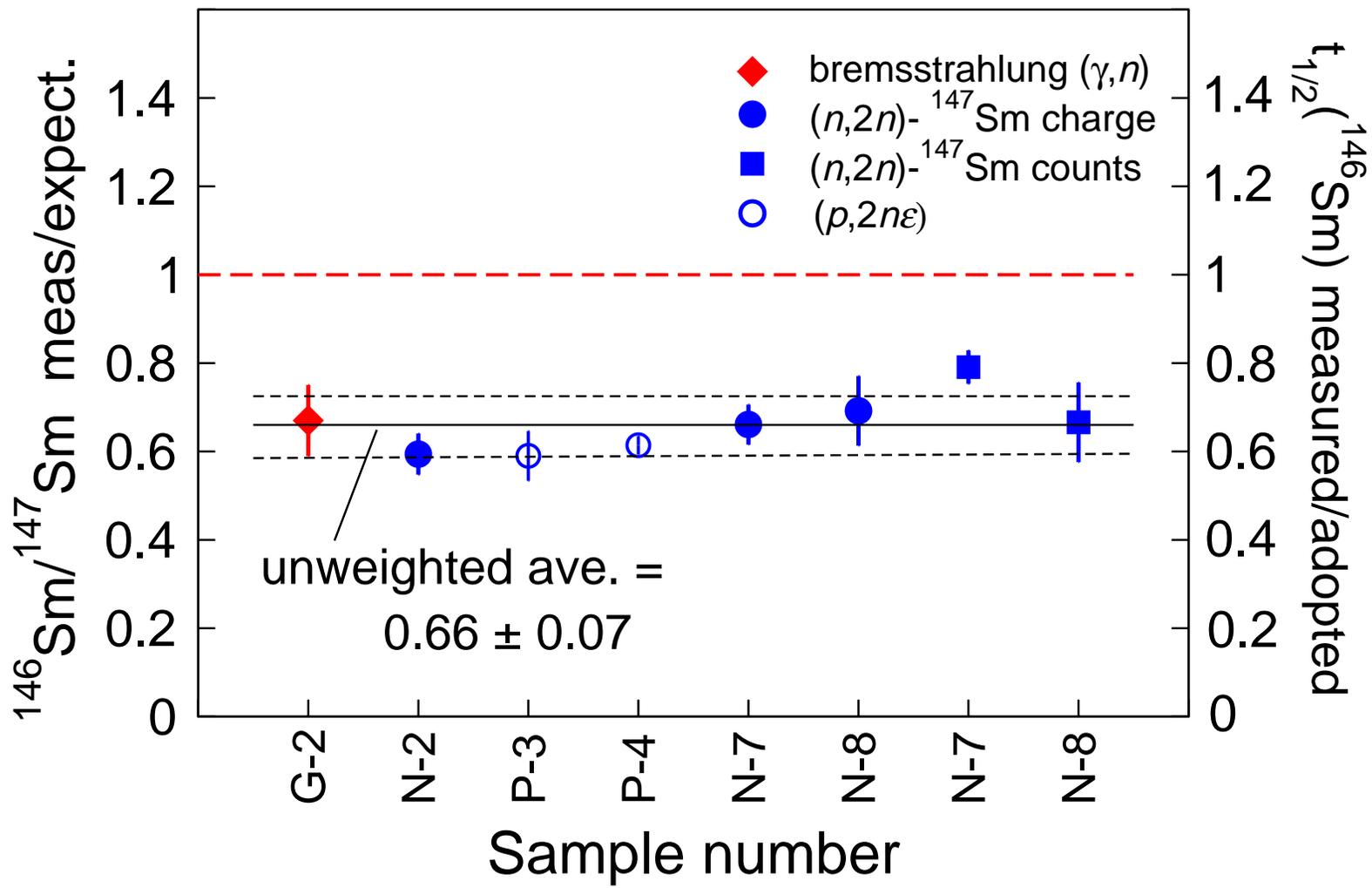

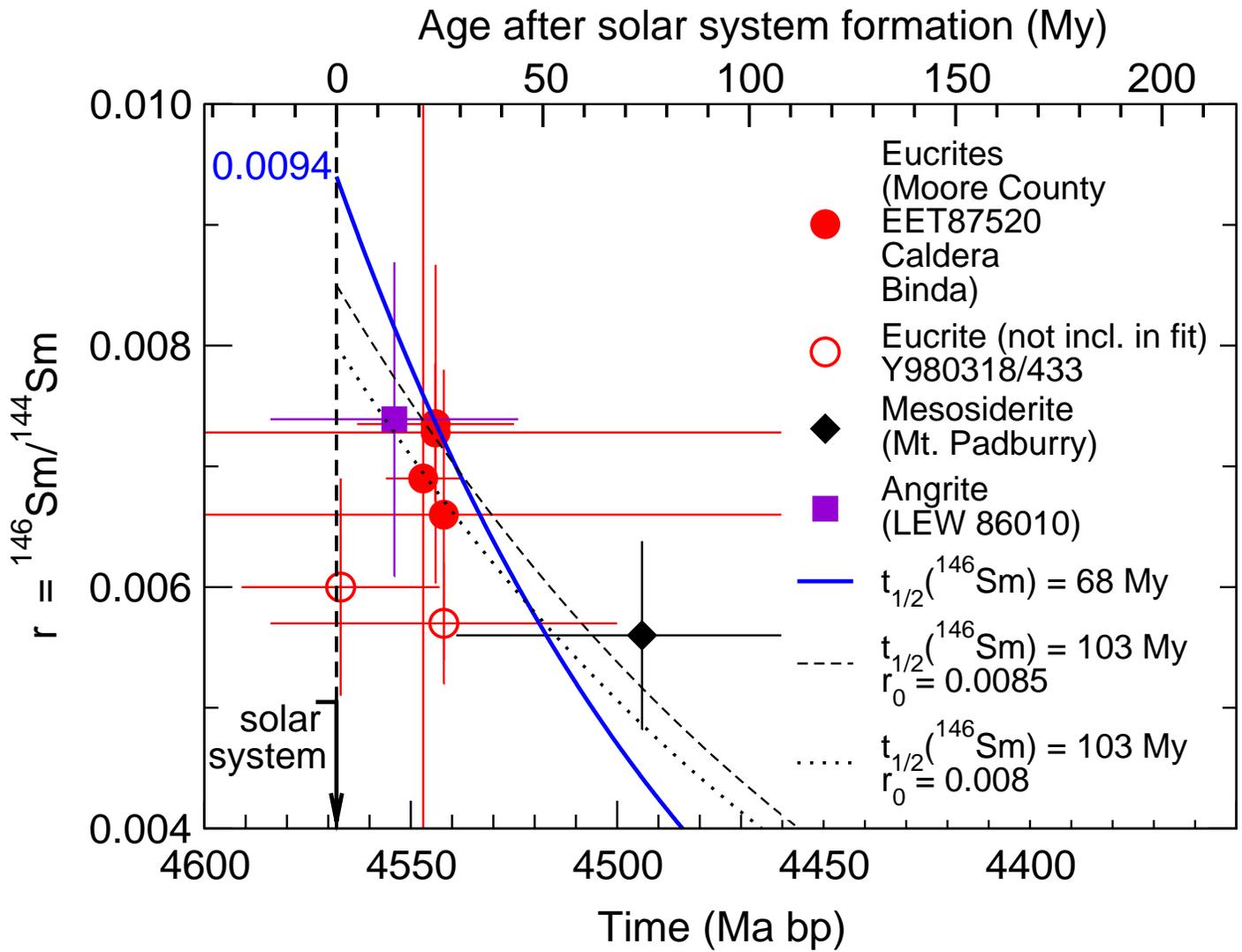

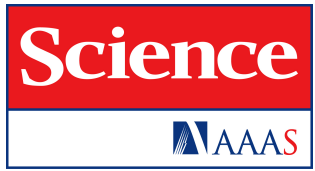

## Supporting Online Material for

### A Shorter $^{146}$Sm Half-Life Measured and Implications for $^{146}$Sm-$^{142}$Nd Chronology in the Solar System

N. Kinoshita, M. Paul, Y. Kashiv, P. Collon, C. M. Deibel, B. DiGiovine, J. P. Greene,
D. J. Henderson, C. L. Jiang, S. T. Marley, T. Nakanishi, R. C. Pardo, K. E. Rehm,
D. Robertson, R. Scott, C. Schmitt, X. D. Tang, R. Vondrasek, A.Yokoyama

correspondence to:  paul@vms.huji.ac.il

**This PDF file includes:**

Materials and Methods
SOM text
Figs. S1 to S6
Tables S1,S2

**Materials and Methods**

Activation of $^{147}$Sm, sample preparation for alpha-activity and AMS measurements

Three independent $^{146}$Sm source materials were produced for the experiments by activating isotopically enriched targets of $^{147}$Sm via the following nuclear reactions: (*i*) $^{147}$Sm($\gamma,n$)$^{146}$Sm (using bremsstrahlung radiation with an end-point energy of 50 MeV from the Electron Linear Accelerator at Tohoku University, Japan); (*ii*) $^{147}$Sm($p,2n\varepsilon$)$^{146}$Sm (using 21 MeV protons from the AVF cyclotron at Osaka University, Japan) and (*iii*) $^{147}$Sm($n,2n$)$^{146}$Sm (using fast neutrons from the Japan Material Testing Reactor, Oarai, Japan). Following the activations, the Sm targets were dissolved and Sm was purified (see protocol of chemical procedure below). In the case of the proton-induced activation, Eu was separated from the Sm target and $^{146}$Eu let to decay to $^{146}$Sm for about three months. $^{146}$Sm decay products and residual Sm were then chemically separated and purified.

Spectroscopic alpha sources (20-100 μg) from the three activations were prepared by precipitation onto Teflon micropore filters (4.9 cm$^2$ filtering area). Alpha activities (Fig. 1, see Table S1) were measured during several months at Kanazawa University (Japan), using a silicon surface-barrier detector of 450 mm$^2$ active area and ~20% efficiency (*34*).

In order to prepare samples for atom ratio measurements, the Sm sources were dissolved and quantitatively diluted with high-purity $^{nat}$Sm in various ratios (Table S1), in order to have samples of typically 10 mg having an effective natural Sm isotopic composition. The solutions were purified of Nd impurities by repeated liquid-chromatography steps using a lanthanide-specific resin (Ln resin, manufactured by Eichrom Ltd.); see protocol below. The samples were eventually reduced to high-purity Sm metal, observed to have higher ionization yields in the Electron Cyclotron Resonance (ECR) ion source (see below) than Sm oxide. This was performed by the following steps for each individual sample: (*i*) Sm hydroxide was precipitated with ammonia solution and evaporated to dryness; (*ii*) ignited to oxide in quartz crucible at 600°C; (*iii*) oxide thoroughly mixed with freshly filed Zr powder and pressed to 3mm diameter pellets; (*iv*) pellets introduced in a tightly fitting cylindrical tube made of 0.1 mm thick Ta having 1 mm diameter orifice; (*v*) sealed tube placed in high-vacuum ($10^{-5}$ Pa) evaporator and resistively heated to ~1300°C; (*vi*) reduced and evaporated metallic Sm collected by distillation on water-cooled Cu collector placed just above the 1-mm orifice. The Sm sample was eventually pressed in a holder made of high-purity Al to be used as sputter cathode in the ECR ion source.

Protocol and chemical procedures

Protocols and chemical procedures used to prepare the activated samples for alpha-activity and atom ratio measurements are shown below as flow charts.

# Chemistry of (p,2nε) sample P-x

```
                target 96% enriched ¹⁴⁷Sm₂O₃ 35 mg
                              │
                              │◄──── 13.5 M HNO₃
                              │
                    evaporated to ~0.1 mL
                    ╱                ╲
                1/2                    1/2
```

- target 96% enriched $^{147}Sm_2O_3$ 35 mg
- ← 13.5 M $HNO_3$
- evaporated to ~0.1 mL
- (left branch, 1/2): same procedure as in right-hand branch → α spectrometry → same dilution procedure as in right branch → AMS sample P-3
- (right branch, 1/2): cation exchange HPLC column (4.6 mm$\phi$ × 250 mm)
  - water 60 mL →
  - 0.3 M α-HIBA solution →
  - → discard
  - → discard (Sm fraction)
  - → Eu + ~30 μg of Sm
- let stand for 3 months
- $^{146}Eu \rightarrow {}^{146}Sm$
  - ← Fe carrier 1 mg
  - ← 12 M HCl
  - ← $NH_3$ water
- precipitate (Fe, Sm)  |  supernatant (α-HIBA removed)   ×2
- precipitate ← 12 M HCl
- Dowex 1X8
  - 12 M HCl →
  - → $Sm^{3+}$
- evaporated to ~0.1 mL

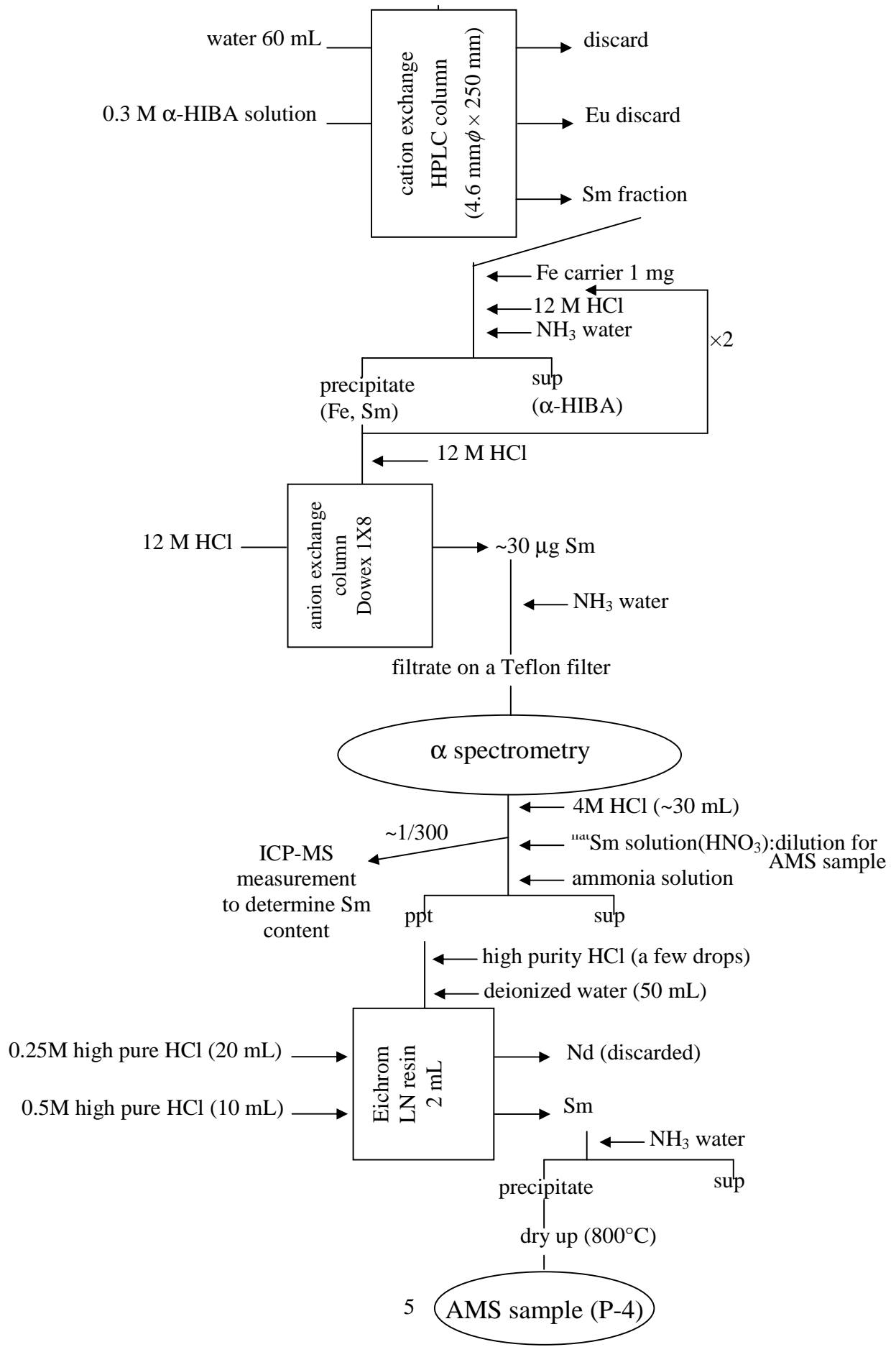

5  AMS sample (P-4)

# Chemistry of (γ,n) sample G-x

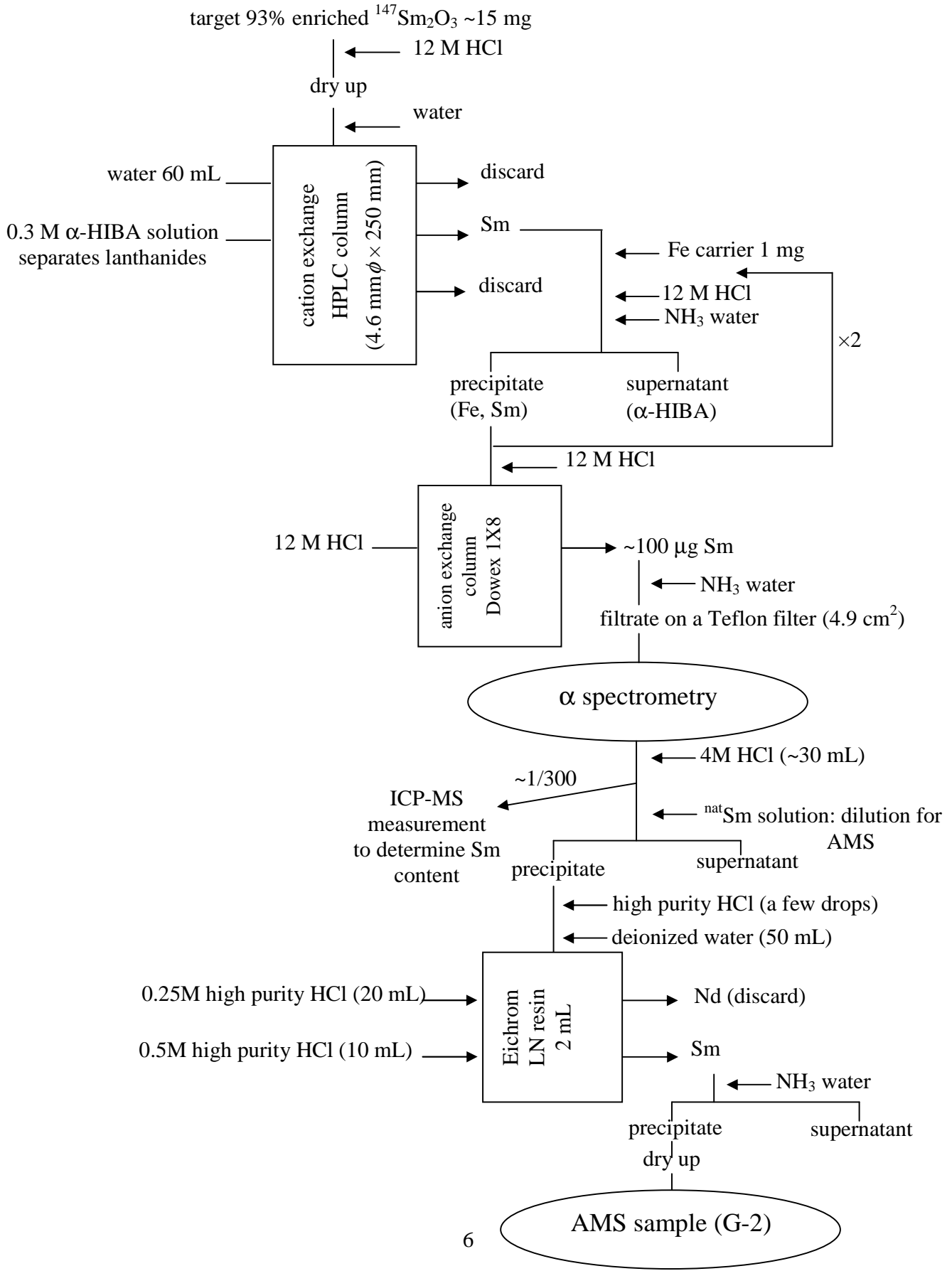

# Chemistry of (n,2n) sample N-x

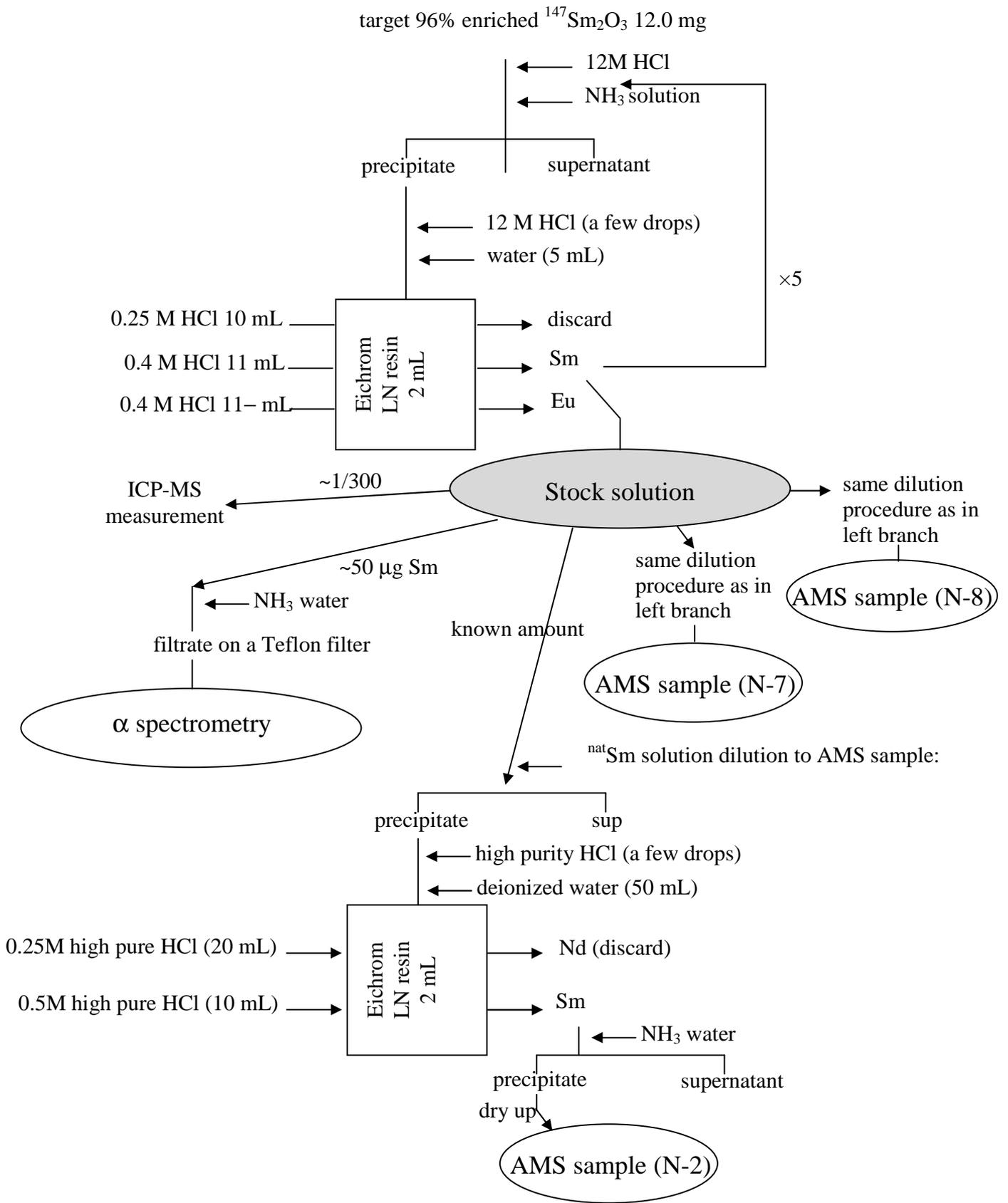

Detection of $^{146}$Sm by accelerator mass spectrometry at the ATLAS facility

Accelerator mass spectrometry (AMS) was used for the determination of $^{146}$Sm/$^{147}$Sm because of the need to discriminate against isobaric $^{146}$Nd originating from residual impurities and structural materials of the ion source. The AMS $^{146}$Sm/$^{147}$Sm measurements were made with the Electron Cyclotron Resonance Ion Source (ECRIS)-ATLAS accelerator facility at Argonne National Laboratory.

Highly-charged positive ions are produced in the plasma chamber of the ECRIS (*35*) by ion sputtering of the Sm metallic sample. In order to reduce background of parasitic ions, the chamber walls were lined with a closely fit cylinder made of quartz transparent to the microwave radiation (~100W at 14 GHz) which ignites and maintains the plasma. $^{146}$Sm$^{22+}$ and $^{147}$Sm$^{22+}$ ions, alternately selected by magnetic separation, were bunched and accelerated to a final energy of 6 MeV/*u* in the superconducting linear accelerator. Acceleration of neighboring isotopes such as $^{146}$Sm and $^{147}$Sm, both in 22+ charge state, requires adjustment of resonator field amplitudes and ion-optical elements by scaling a master tune according to the *m/q* ratios of the atomic mass and charge state of the ions, in order to conserve an exact velocity profile. The accelerator system acts then as a high-energy accelerator mass spectrometer and its abundance sensitivity (defined as the abundance ratio of two resolvable neighboring isotopes) before the detection system was shown (*36*) to be ~5×10$^{-12}$ for Pb isotopes, separating in the present experiment $^{146}$Sm from $^{147}$Sm. Molecular ions are totally dissociated in the ECR plasma and any molecular ion re-formed at a later stage cannot be injected at the magnetic rigidity corresponding to the high charge-state ions selected for acceleration.

Isobaric separation of $^{146}$Sm and residual $^{146}$Nd (($m/q$)/$\delta$($m/q$)~2.4×10$^6$), beyond the separation power of the accelerator, is achieved in a gas-filled magnetic spectrograph (*27*). The ions are then physically separated due to their (Z-dependent) mean charge state in the gas, unambiguously identified and counted by measurement of their position and differential energy loss in a focal-plane detector (Fig. 1 right panel). For a quantitative determination of the atom ratio, the count rate of $^{146}$Sm$^{22+}$ ions in the focal-plane detector was normalized to the charge current (of the order of several nA) of the $^{147}$Sm$^{22+}$ beam, transported after proper scaling of the accelerator to electron-suppressed Faraday cup positioned before the spectrograph entrance. The fraction of ions produced by the ECRIS in the 22+ charge state is considered identical for the two isotopes. The atom ratio for each measured sample is listed in Table S1 and is shown (Fig. 2 and Fig. S3), divided by the expected atom ratio calculated using its alpha activity, its dilution factor (Table S1) and the currently adopted half-life value of $^{146}$Sm (103 Ma, (*22,23*)). The double ratio, equivalent to the ratio of our measured half-life to that currently adopted is 0.66 ± 0.07 (unweighted average of the different samples and standard deviation). The corresponding $^{146}$Sm half-life value is 68 ± 7 Ma. The standard deviation accounts for random errors in ion-counting, alpha activity, dilution and fluctuations in ion transmission of the accelerator system (Figs. 2, S2, S3, Table S1). In order to estimate the effective uncertainty associated with these fluctuations, difficult to control in the accelerator, we performed repeat measurements of each sample and conservatively take the unweighted standard deviation of the measurements as the uncertainty in each set of data points. The systematic uncertainty in our determination of the $^{146}$Sm half-life introduced by the uncertainty in the $^{147}$Sm half-life is negligible, when using the high precision value (107.0 ± 0.9 Ga) (*24*); see Table S2 for a compilation of available data on $^{147}$Sm half-life. The reliability of the *m/q* scaling procedure used to transport alternately $^{146}$Sm$^{22+}$ or $^{147}$Sm$^{22+}$ from the ECRIS to the detection setup, was checked by measuring the ratio of charge currents for $^{147}Sm^{22+}$ and $^{152}Sm^{22,23+}$ (Fig. S2). The measured $^{152}Sm/^{147}Sm$ (1.8±0.2), after correcting when necessary for the 23+/22+ population ratio (Figs. S1, S2), is consistent with the natural abundance ratio 1.785 ± 0.024 (derived from (*37*)), showing that the average transmission at different *m/q* settings is constant.

The measurements of sample N-8 were less reproducible than others, possibly due to lower quality of sample N-8 or lower performance of the accelerator system during the N-8 measurement, resulting in less stable ion transmission. This is observed in the increased scatter of the $^{152}Sm/^{147}Sm$ and $^{146}Sm/^{147}Sm$ ratios of N-8 in Figs. S2 and S3. It is also expressed by the larger random error in the N-8 double ratio (Fig. 2, Table S1) than for other samples. The poorer stability of N-8 led us to take a larger set of repeat measurements of its atom ratios (Figs. S2, S3) in order to obtain a valid result. The random errors assigned to the N-8 data points in Figs. S2 and S3 are, however, statistically consistent with the scatter due to the fluctuations in ion transmission.

For two of the samples (N-7 and N-8), the atom ratios were also measured by alternately counting $^{146}Sm^{22+}$ (unattenuated) and $^{147}Sm^{22+}$ (attenuated) in the focal plane detector. The measurement method is described in detail in (*29*): quantitative attenuation of 1:10$^5$, required for the abundant $^{147}Sm^{22+}$ ions, is obtained with the 12 MHz pulsed and ns-bunched ATLAS beam by chopping ion pulses with an RF sweeper in a (digitally measured) ratio of 1:10$^5$. As shown in (29), the measured atom ratios (Fig. 2) are consistent with those described above and confirm that no unaccounted ion losses occur between the Faraday cup and focal plane detector. The ion spectra accumulated for attenuated beams of $^{147}Sm$ and $^{152}Sm$ (Fig. 2 in *29*) show also that the respective beams are isotopically pure.

**SOM Text**

Sm dilution (P-3, P-4, G-2, N-2,7,8)

**Index**
1. Preparation of Sm solution for dilution and ICP-MS measurement
2. P-3
3. P-4
4. G-2
5. N-2

**Preparation of Sm solution for dilution and ICP-MS measurement**

1. A weighing bottle was dried in a vacuum oven (130 $^oC$) for a couple of hours, the bottle was put in a desiccator for 30 min, and then weight of the bottle was measured. This procedure was repeated totally 3 times.

    Weight: 15.031 g → 15.031 g → 15.031 g   Average: 15.031 g

2. Approximately 1.5 g of natural $Sm_2O_3$ (Nakalai Tesque Co., purity: 99.9%) was taken in the weighing bottle. Then, bottle was dried in a vacuum oven (130 $^oC$) for a couple of hours, the bottle was put in a desiccator for 30 min, and then weight of the bottle was measured. This procedure was also repeated 3 times.

    Weight: 16.288 g → 16.288 g → 16.288 g   Average: 16.288 g
    Weight of $Sm_2O_3$: 16.288 − 15.031 = $\boxed{1.257 \text{ g}}$

3. The $Sm_2O_3$ was dissolved in ~10 mL of c.$HNO_3$ and then diluted to $\boxed{50.003 \text{ g}}$ with deionized water.

$$\text{Sm concentration is:} 1.257 \times \frac{300.36}{348.72} \times 0.999 \div 50.003 = 21.63$$

$$= \boxed{21.6 \text{ mg }^{nat}\text{Sm/g-soln.}}$$

   This Sm solution was used for the following dilution (P-3, P-4, G-2, N-2).

4. Preparation of Sm solution for ICP-MS measurement.
    Sm-1: 55.9 mg of 21.6 mg $^{nat}$Sm/g-soln. was diluted to 58.7 g → 20.6 μg $^{nat}$Sm/g
    Sm-2: 258.9 mg of Sm-1(20.6 μg/g) was diluted to 65.31 g → 81.7 ng $^{nat}$Sm/g

Sm-3: 52.0 mg of Sm-1(20.6 µg/g) was diluted to 60.59 g → 17.7 ng $^{nat}$Sm/g

Sm-4: 5.194 g of Sm-2(81.7 µg/g) was diluted to 58.87 g → 7.21 ng $^{nat}$Sm/g

Sm-5: 2.587 g of Sm-2(81.7 µg/g) was diluted to 61.50 g → 3.44 ng $^{nat}$Sm/g

Sm-6: 511.1 mg of Sm-2(81.7 µg/g) was diluted to 59.97 g → 0.696 ng $^{nat}$Sm/g

Sm-7: 52.7 mg of Sm-2(81.7 µg/g) was diluted to 62.89 g → 0.0333 ng $^{nat}$Sm/g

Those Sm solutions (Sm-3 - Sm-7) were used for ICP-MS measurement.

## P-3

### Measurement of $^{147}$Sm content

1. Alpha measurement sample, which Sm(OH)$_3$ was mounted on a Teflon filter, was soaked in ~4M HCl.

    Alpha activity ratio $^{146}$Sm/$^{147}$Sm: $(3.01 \pm 0.06) \times 10^{-1}$

2. The filter was removed and weight of the solution was measured.

    The weight: 36.0 g

3. 51.3 mg (=50 µL) of the solution was diluted to 49.8g ($\cong$ 50 mL) with ~1M HNO$_3$ for ICP-MS measurement.

4. $^{147}$Sm content in the solution was determined by ICP-MS.

| $^{nat}$Sm concentration (ng/g-soln.) | $^{147}$Sm concentration (atoms/g-soln.) | count rate (cps) |
|---|---|---|
| 0.0333 | 2.00E+10 | 210.2 ± 19.3 |
| 0.696 | 4.18E+11 | 2067.8 ± 13.6 |
| 3.44 | 2.06E+12 | 10018.4 ± 94.2 |
| 7.21 | 4.33E+12 | 20550.6 ± 137.7 |
| 17.6 | 1.06E+13 | 49580.0 ± 738.7 |
| P-3 sample | | 14351.2 ± 196.6 |

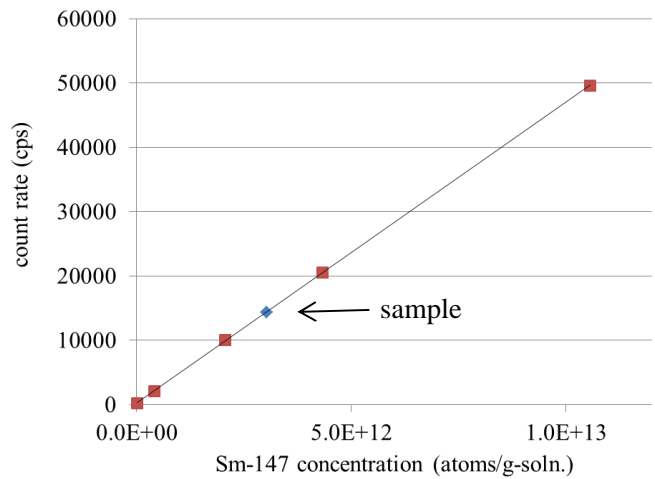

Concentration of activated $^{147}$Sm solution is : $\dfrac{(14351.2 \pm 196.6) - 202.8}{4.684 \times 10^{-9}} \times \dfrac{49.8}{51.3 \times 10^{-3}}$

$= \boxed{(2.92 \pm 0.06) \times 10^{15}}$ $^{147}$Sm atoms/g

**Dilution and purification for AMS measurement**

5. 36.6 g of activated $^{147}$Sm solution was mixed with 4.45 g (= 4 mL, $\boxed{5.78 \times 10^{19}\ ^{147}\text{Sm atoms}}$) of $^{nat}$Sm solution (21.6 mg $^{nat}$Sm/g-soln.).
6. The solution was mixed well and then Sm was precipitated with ammonia water.
7. The Sm precipitate was dissolved in a few drops of HCl and diluted to ~50 mL with deionized water.
8. 1 mL of the solution was loaded on each Ln resin column. 20 mL of 0.25M HCl was passed through the column. Subsequently, 0.50 M HCl was passed through the column. Sm comes out with 0.50 M HCl. Sm was precipitated with ammonia water. The precipitate was heated in an oven (450 $^{\circ}$C, 7 hours). The precipitate was shipped to Argonne.

**Dilution rate** is : $\dfrac{(2.92 \pm 0.06) \times 10^{15} \times 36.6 + 5.78 \times 10^{19}}{(2.92 \pm 0.06) \times 10^{15} \times 36.6} = \boxed{551 \pm 8}$

# P-4

## Measurement of $^{147}$Sm content

1. Alpha measurement sample, which Sm(OH)$_3$ was mounted on a Teflon filter, was soaked in ~4M HCl.

   Alpha activity ratio $^{146}$Sm/$^{147}$Sm: $(4.58 \pm 0.10) \times 10^{-1}$

2. The filter was removed and weight of the solution was measured.
   The weight: $\boxed{37.7\text{g}}$
3. $\boxed{51.9\text{ mg}}$ (=50 μL) of the solution was diluted to $\boxed{51.7\text{g}}$ ($\cong$ 50 mL) with ~1M HNO$_3$ for ICP-MS measurement.

4. $^{147}$Sm content in the solution was determined by ICP-MS.

| $^{nat}$Sm concentration (ng/g-soln.) | $^{147}$Sm concentration (atoms/g-soln.) | count rate (cps) |
|---|---|---|
| 0.0333 | 2.00E+10 | 210.2 ± 19.3 |
| 0.696 | 4.18E+11 | 2067.8 ± 13.6 |
| 3.44 | 2.06E+12 | 10018.4 ± 94.2 |
| 7.21 | 4.33E+12 | 20550.6 ± 137.7 |
| 17.6 | 1.06E+13 | 49580.0 ± 738.7 |
| P-4 sample | | 8252.2 ± 90.0 |

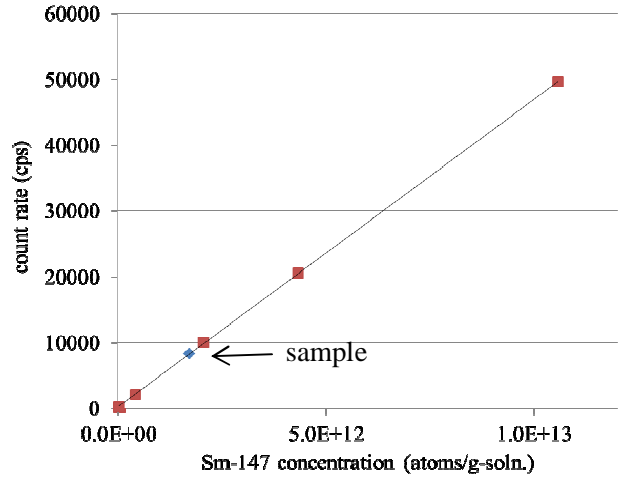

Concentration of activated $^{147}$Sm solution is

$$\frac{(8252.2 \pm 90.0) - 202.8}{4.684 \times 10^{-9}} \times \frac{51.7}{51.9 \times 10^{-3}} = \boxed{(1.71 \pm 0.02) \times 10^{15} \ ^{147}\text{Sm atoms/g}}$$

**Dilution and purification for AMS measurement**

5. 37.7 g of activated $^{147}$Sm solution was mixed with 4.44 g (= 4 mL, $5.76 \times 10^{19}$ $^{147}$Sm atoms) of $^{nat}$Sm solution (21.6 mg $^{nat}$Sm/g-soln.).
6. The solution was mixed well and then Sm was precipitated with ammonia water.
7. The Sm precipitate was dissolved in a few drops of HCl and diluted to ~50 mL with deionized water.
8. 1 mL of the solution was loaded on each Ln resin column. 20 mL of 0.25M HCl was passed through the column. Subsequently, 0.50 M HCl was passed through the column. Sm comes out with 0.50 M HCl. Sm was precipitated with ammonia water. The precipitate was heated in an oven (450 °C, 7 hours). The precipitate was shipped to Argonne.

**Dilution rate** is : $\dfrac{(1.71 \pm 0.02) \times 10^{15} \times 37.7 + 5.76 \times 10^{19}}{(1.71 \pm 0.02) \times 10^{15} \times 37.7} = \boxed{894 \pm 10}$

# G-2

**Measurement of $^{147}$Sm content**

1. Alpha measurement sample, which Sm(OH)$_3$ was mounted on a Teflon filter, was soaked in ~4M HCl.

    Alpha activity ratio $^{146}$Sm/$^{147}$Sm: $(4.04 \pm 0.14) \times 10^{-3}$

2. The filter was removed and weight of the solution was measured.

    The weight: 38.7g

3. 51.2 mg (=50 μL) of the solution was diluted to 51.9g ($\cong$ 50 mL) with ~1M HNO$_3$ for ICP-MS measurement.

4. $^{147}$Sm content in the solution was determined by ICP-MS.

| $^{nat}$Sm concentration (ng/g-soln.) | $^{147}$Sm concentration (atoms/g-soln.) | count rate (cps) |
|---|---|---|
| 0.0333 | 2.00E+10 | 210.2 ± 19.3 |
| 0.696 | 4.18E+11 | 2067.8 ± 13.6 |
| 3.44 | 2.06E+12 | 10018.4 ± 94.2 |
| 7.21 | 4.33E+12 | 20550.6 ± 137.7 |
| 17.6 | 1.06E+13 | 49580.0 ± 738.7 |
| G-2 sample | | 50862.0 ± 213.6 |

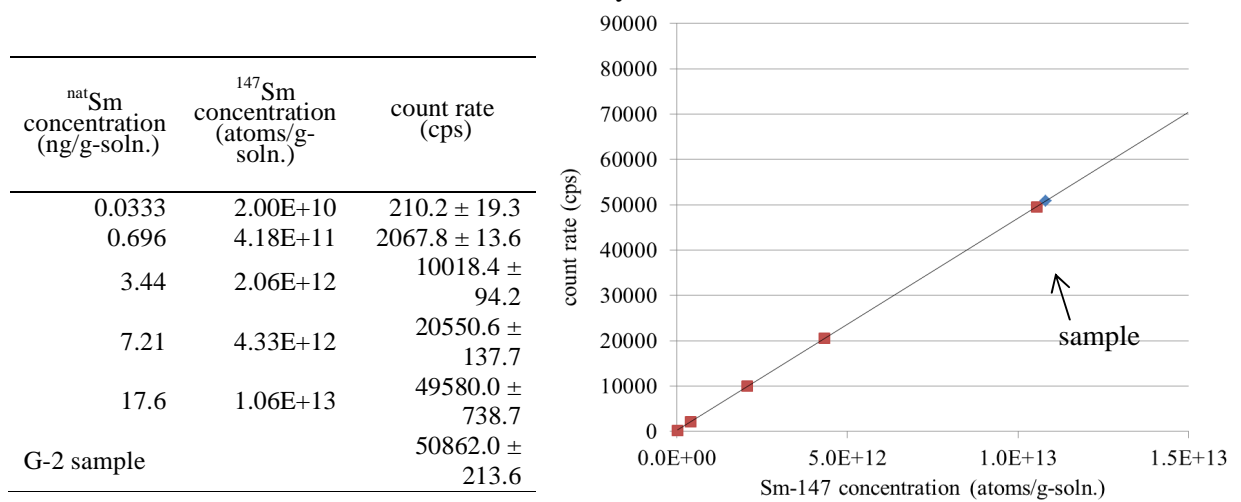

Concentration of activated $^{147}$Sm solution is: $\dfrac{(50862.0 \pm 213.6) - 202.8}{4.684 \times 10^{-9}} \times \dfrac{51.9}{51.2 \times 10^{-3}} =$

$= (1.09 \pm 0.01) \times 10^{17}\ ^{147}$Sm atoms/g

**Dilution and purification for AMS measurement**

5. 38.7 g of activated $^{147}$Sm solution was mixed with 4.46 g (= 4 mL, $5.78 \times 10^{19}$ $^{147}$Sm atoms) of $^{nat}$Sm solution (21.6 mg $^{nat}$Sm/g-soln.).

6. The solution was mixed well and then Sm was precipitated with ammonia water.

7. The Sm precipitate was dissolved in a few drops of HCl and diluted to ~50 mL with deionized water.

8. 1 mL of the solution was loaded on each Ln resin column. 20 mL of 0.25M HCl was passed through the column. Subsequently, 0.50 M HCl was passed through the column. Sm comes out with 0.50 M HCl. Sm was precipitated with ammonia water. The precipitate was heated in an oven (450 °C, 7 hours). The precipitate was shipped to Argonne.

**Dilution rate** is : $\dfrac{(1.09 \pm 0.01) \times 10^{16} \times 38.7 + 5.78 \times 10^{19}}{(1.09 \pm 0.01) \times 10^{16} \times 38.7} = \boxed{137 \pm 1}$

## N-2, N-7, N-8
### Measurement of $^{147}$Sm content

1. $\boxed{1.09 \text{ g}}$ of activated $^{147}$Sm stock solution was diluted to $\boxed{34.3 \text{ g}}$ with ~1M HNO$_3$.
   Alpha activity ratio $^{146}$Sm/$^{147}$Sm: $(6.27 \pm 0.14) \times 10^{-3}$
2. In addition, $\boxed{51.8 \text{ mg}}$ (=50 µl) of the diluted solution was diluted to $\boxed{50.8\text{g}}$ ($\cong$ 50 mL) with ~1M HNO$_3$ for ICP-MS measurement.
3. $^{147}$Sm content in the solution was determined by ICP-MS.

| $^{nat}$Sm concentration (ng/g-soln.) | $^{147}$Sm concentration (atoms/g-soln.) | count rate (cps) |
|---|---|---|
| 0.0333 | 2.00E+10 | 210.2 ± 19.3 |
| 0.696 | 4.18E+11 | 2067.8 ± 13.6 |
| 3.44 | 2.06E+12 | 10018.4 ± 94.2 |
| 7.21 | 4.33E+12 | 20550.6 ± 137.7 |
| 17.6 | 1.06E+13 | 49580.0 ± 738.7 |
| N-2 sample | | 80302.0 ± 465.8 |

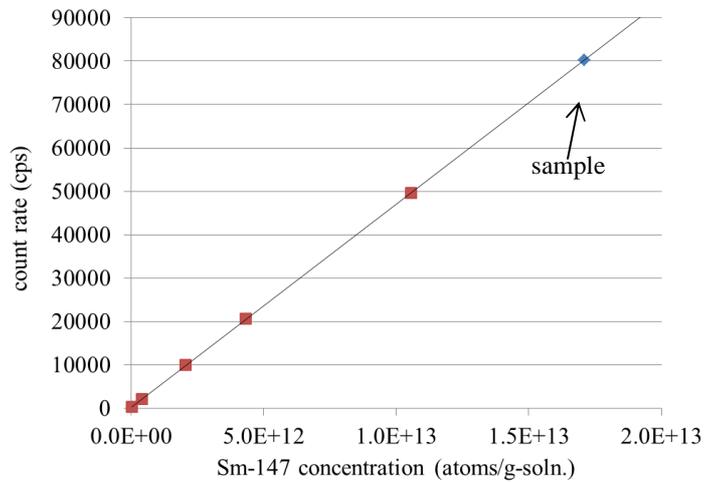

Concentration of activated $^{147}$Sm solution is:

$\dfrac{(80302 \pm 465.8) - 202.8}{4.684 \times 10^{-9}} \times \dfrac{50.8}{51.8 \times 10^{-3}} \times \dfrac{34.3}{1.09} = \boxed{(5.28 \pm 0.03) \times 10^{17}}$ $^{147}$Sm atoms/g

**Dilution and purification for AMS measurement (N-2)**

4. 1.08g of activated $^{147}$Sm solution was mixed with 4.43 g (= 4 mL, $5.75 \times 10^{19}$ $^{147}$Sm atoms) of $^{nat}$Sm solution (21.6 mg $^{nat}$Sm/g-soln.).
5. The solution was mixed well and then Sm was precipitated with ammonia water.
6. The Sm precipitate was dissolved in a few drops of HCl and diluted to ~50 mL with deionized water.
7. 1 mL of the solution was loaded on each Ln resin column. 20 mL of 0.25M HCl was passed through the column. Subsequently, 0.50 M HCl was passed through the column. Sm comes out with 0.50 M HCl. Sm was precipitated with ammonia water. The precipitate was heated in an oven (450 $^o$C, 7 hours). The precipitate was shipped to Argonne.

**Dilution rate** is: $\dfrac{(5.28 \pm 0.03) \times 10^{17} \times 1.08 + 5.75 \times 10^{19}}{(5.28 \pm 0.03) \times 10^{17} \times 1.08} = \boxed{102 \pm 1}$

**Dilution and purification for AMS measurement (N-7)**

4. 2.013g of activated $^{147}$Sm solution was mixed with 8.311 g (= 8 mL, $3.42 \times 10^{19}$ $^{147}$Sm atoms) of $^{nat}$Sm solution (6.86 mg $^{nat}$Sm/g-soln.).
5. The solution was mixed well and then Sm was precipitated with ammonia water.
6. The Sm precipitate was dissolved in a few drops of HCl and diluted to ~50 mL with deionized water.
7. 1 mL of the solution was loaded on each Ln resin column. 20 mL of 0.25M HCl was passed through the column. Subsequently, 0.50 M HCl was passed through the column. Sm comes out with 0.50 M HCl. Sm was precipitated with ammonia water. The precipitate was heated in an oven (450 $^o$C, 7 hours). The precipitate was shipped to Argonne.

**Dilution rate** is: $\dfrac{(5.28 \pm 0.03) \times 10^{17} \times 2.013 + 3.42 \times 10^{19}}{(5.28 \pm 0.03) \times 10^{17} \times 2.013} = \boxed{33.2 \pm 0.2}$

**Dilution and purification for AMS measurement (N-8)**

4. 1.020g of activated $^{147}$Sm solution was mixed with 3.164g (= 3 mL, $(4.94\pm0.19)\times10^{19}$ $^{147}$Sm atoms) of $^{nat}$Sm solution ($26.0\pm1.0$ mg $^{nat}$Sm/g-soln.).
5. The solution was mixed well and then Sm was precipitated with ammonia water.
6. The Sm precipitate was dissolved in a few drops of HCl and diluted to ~50 mL with deionized water.
7. 1 mL of the solution was loaded on each Ln resin column. 20 mL of 0.25M HCl was passed through the column. Subsequently, 0.50 M HCl was passed through the column. Sm comes out with 0.50 M HCl. Sm was precipitated with ammonia water. The precipitate was heated in an oven (450 °C, 7 hours). The precipitate was shipped to Argonne.

**Dilution rate** is : $\dfrac{(5.28\pm0.03)\times10^{17}\times1.020 + (4.94\pm0.19)\times10^{19}}{(5.28\pm0.03)\times10^{17}\times1.020}$ = $\boxed{92.8 \pm 5.0}$

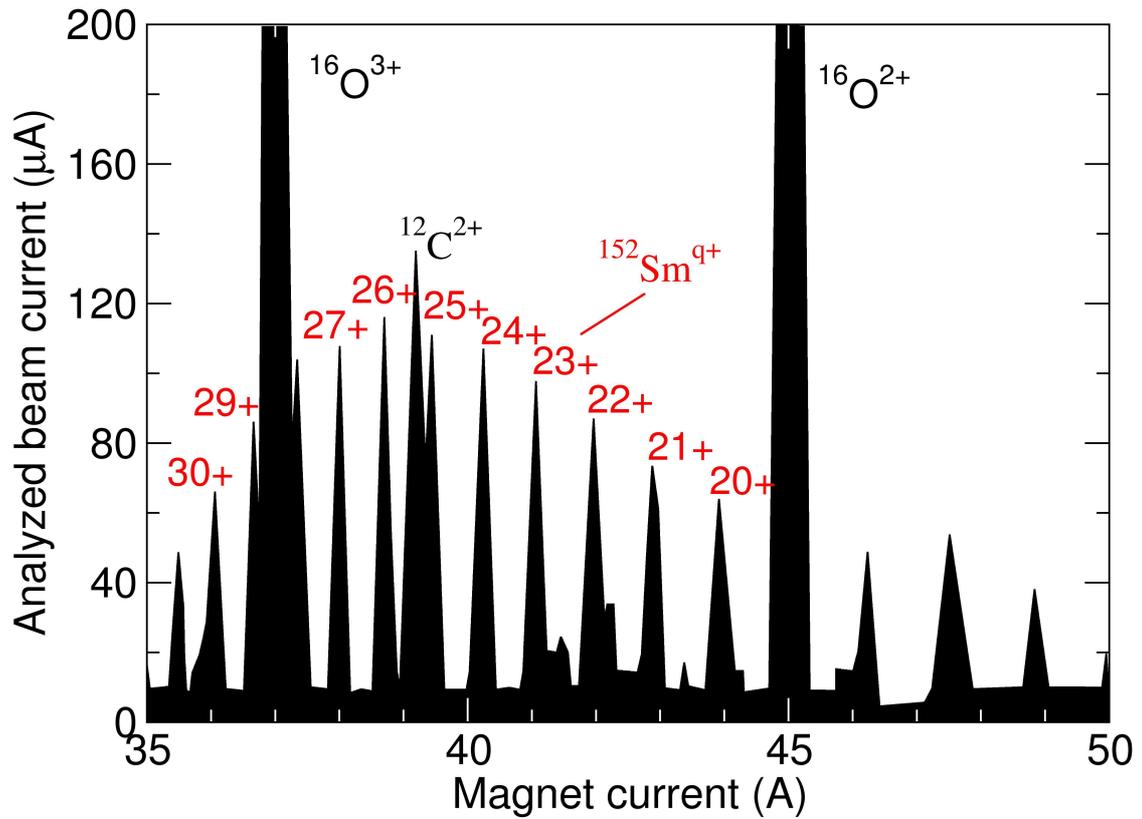

**Fig. S1.**
Charge state distribution measured after the first analysis magnet following the ECR ion source for a metallic sample of enriched $^{152}$Sm. The magnet current on the horizontal axis determines the magnetic field of the injection mass spectrometer. The analyzed beam current (vertical axis) was measured in a Faraday cup at the image position of the analysis magnet. Charge state 22+ for $^{147}$Sm and 22+, 23+ for $^{152}$Sm were used in the experiment and the charge state distribution shown served for normalization.

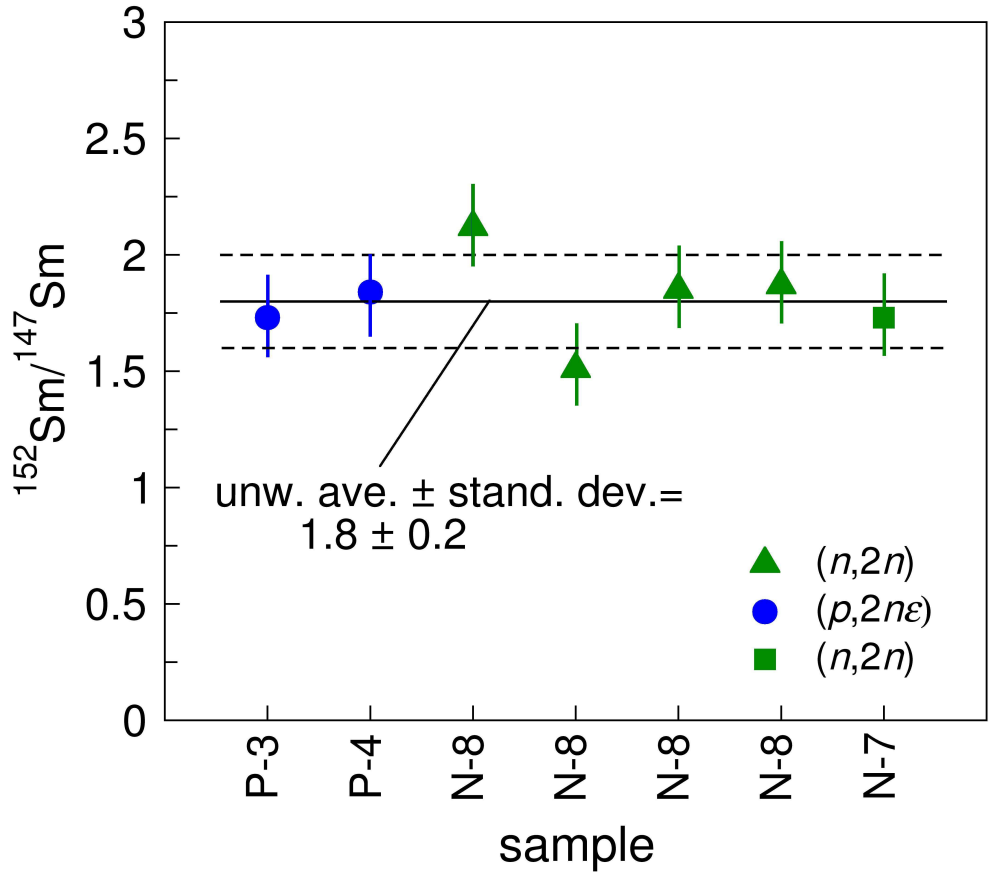

**Fig. S2**

Isotopic ratio $^{152}Sm/^{147}Sm$ measured during our experiment using the same accelerator scaling procedure as used for the ratio $^{146}Sm/^{147}Sm$. The unweighted average and standard deviation value, 1.8 ± 0.2 shown in the figure, is consistent with the natural ratio 1.785 ± 0.024 (derived from (*37*)). Sample N-8 showed larger fluctuations than other samples and additional measurements were made. Error bars represent random errors estimated in ion transmission through the accelerator; the standard deviation of the data points (0.2) is consistent with these random errors.

**Fig. S3**

Individual (repeat) determinations of the double ratio of the $^{146}$Sm/$^{147}$Sm atom ratio measured by AMS in individual repeat measurements of various samples to that expected in the same sample from its $^{146}$Sm/$^{147}$Sm α activity ratio using the $^{146,147}$Sm currently adopted half-lives. The double ratio is equivalent to the ratio of the presently measured $^{146}$Sm half-life to that adopted in the literature. Error bars represent statistical errors in alpha activity and ion counting, random errors in dilution ratios and ion transmission. The AMS atom ratios are measured relative to the charge current of $^{147}$Sm$^{22+}$ measured in an electron-suppressed Faraday Cup in front of the gas-filled magnetic spectrograph. Fluctuations observed for sample N-8 are larger than for other samples but consistent with the random errors assigned (see also Fig. S2). The unweighted average of the individual points is slightly different from the average (0.66) in Fig. 2 because there data of each sample were grouped and averaged. The uncertainty shown is the standard deviation of the data points (unweighted).

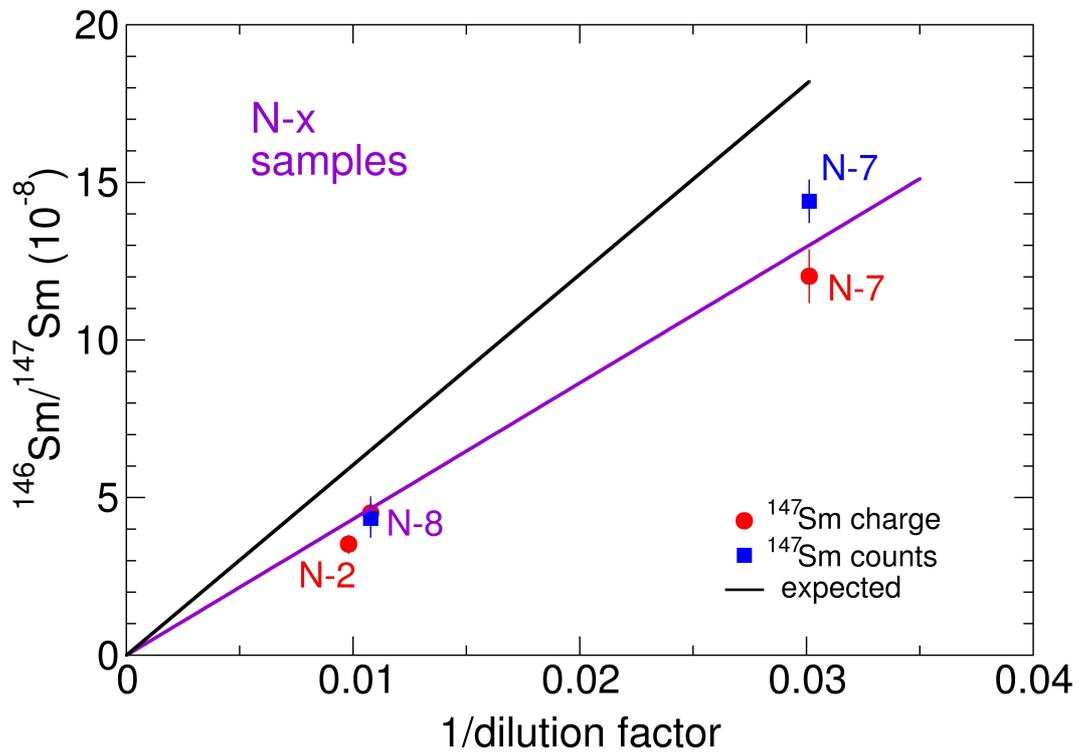

**Fig. S4**

Dilution plot for samples N-x (Table S1) prepared by dilution from a common stock solution of the neutron activated ($^{147}$Sm($n,2n$)$^{146}$Sm) material. The horizontal axis represents the inverse of the dilution factor and the vertical axis the atom ratio $^{146}$Sm/$^{147}$Sm measured by AMS after dilution. Solid dots correspond to normalization by $^{147}$Sm$^{22+}$ charge and solid squares by $^{147}$Sm counts (see SOM Materials and methods). The lower solid line represents a best linear fit of the data points whose slope is proportional to the measured $^{146}$Sm half-life (68 ± 7 Ma). The upper solid line is the dilution line for the N-x samples expected from the α activity measured and the currently adopted $^{146}$Sm half-life value (103 Ma).

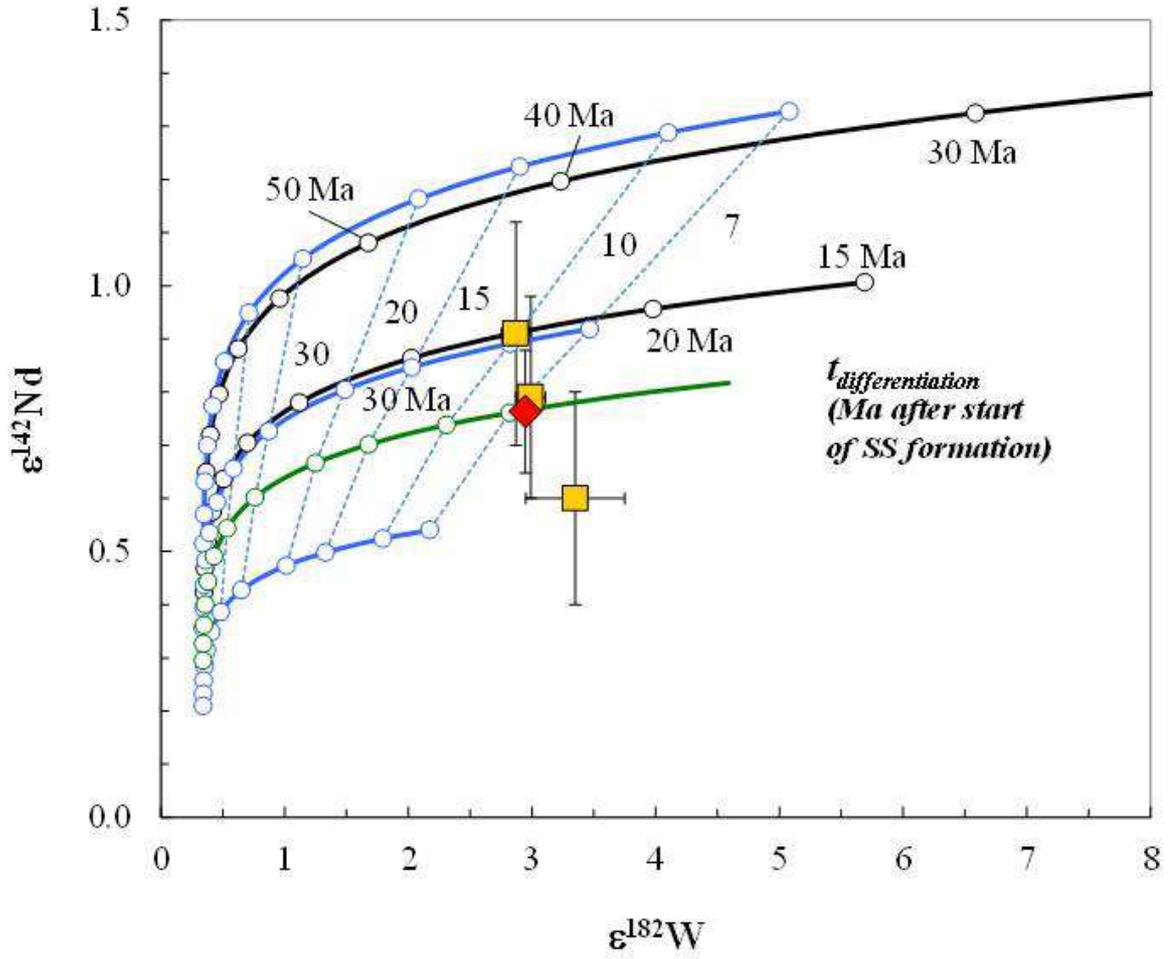

**Fig. S5**

Reinterpretation of the differentiation age and $^{147}$Sm/$^{144}$Nd ratio of the depleted Martian mantle source of nakhlite meteorites (Foley *et al.* (*16*) and references therein, compare with Fig.5 in (*16*)), using the values derived in our work for the $^{146}$Sm half-life $t^{146}_{1/2}$ = 68 Ma and solar system initial ratio $r_0$ = ($^{146}$Sm/$^{144}$Sm)$_0$ = 0.0094. The axes represent the deviation in $^{142}$Nd isotopic abundance from CHUR in units of $10^{-4}$ ($\varepsilon^{142}Nd = \left(\frac{\left(^{142}Nd/^{144}Nd\right)^{DM}}{\left(^{142}Nd/^{144}Nd\right)^{CHUR}} - 1\right) \times 10^4$, where DM stands for depleted mantle) plotted against the deviation of $^{182}$W isotopic abundance ($\varepsilon^{182}W = \left(\frac{\left(^{182}W/^{183}W\right)^{DM}}{\left(^{182}W/^{183}W\right)^{standard}} - 1\right) \times 10^4$). Yellow squares are the individual nakhlites measured and the red diamond is their mean (*16*).

Black curves are models calculated using eqs. 5-6 of Foley *et al*. (*16*) (with the new $t_{1/2}^{146}$ and $r_0$ values): upper curve is a majorite-bearing deep mantle (MDM) source, $^{147}$Sm/$^{144}$Nd ~0.303 and $^{180}$Hf/$^{183}$W ~126; lower curve is garnet-bearing shallow mantle (GSM) source, $^{147}$Sm/$^{144}$Nd ~0.266 and $^{180}$Hf/$^{183}$W ~43. Open circles on the black curves have ages (after start of SS formation) labeled with "Ma". Blue curves are depleted mantle models by Kleine *et al*. (*38*): upper curve is a source with $^{147}$Sm/$^{144}$Nd = 0.281 and $^{180}$Hf/$^{183}$W = 27, middle curve is a source with $^{147}$Sm/$^{144}$Nd = 0.255 and $^{180}$Hf/$^{183}$W = 22, lower curve is a source with $^{147}$Sm/$^{144}$Nd = 0.231 and $^{180}$Hf/$^{183}$W = 18. Labeled dashed lines connect points of same age (in Ma) on blue curves. The mean data point (red diamond) is best matched by a composition $^{147}$Sm/$^{144}$Nd = 0.245 and $^{180}$Hf/$^{183}$W = 20 (green curve). This composition is less depleted than that inferred in Foley *et al*. (*16*), where the data were best explained by $^{147}$Sm/$^{144}$Nd = 0.255-0.266 and $^{180}$Hf/$^{183}$W = 22-43. In addition, the reinterpretation of the data favors the younger part of the 8-25 Ma age range inferred in the original paper. The young age may be in line with the recent finding that Mars accreted within ~ 4 Ma (*33*).

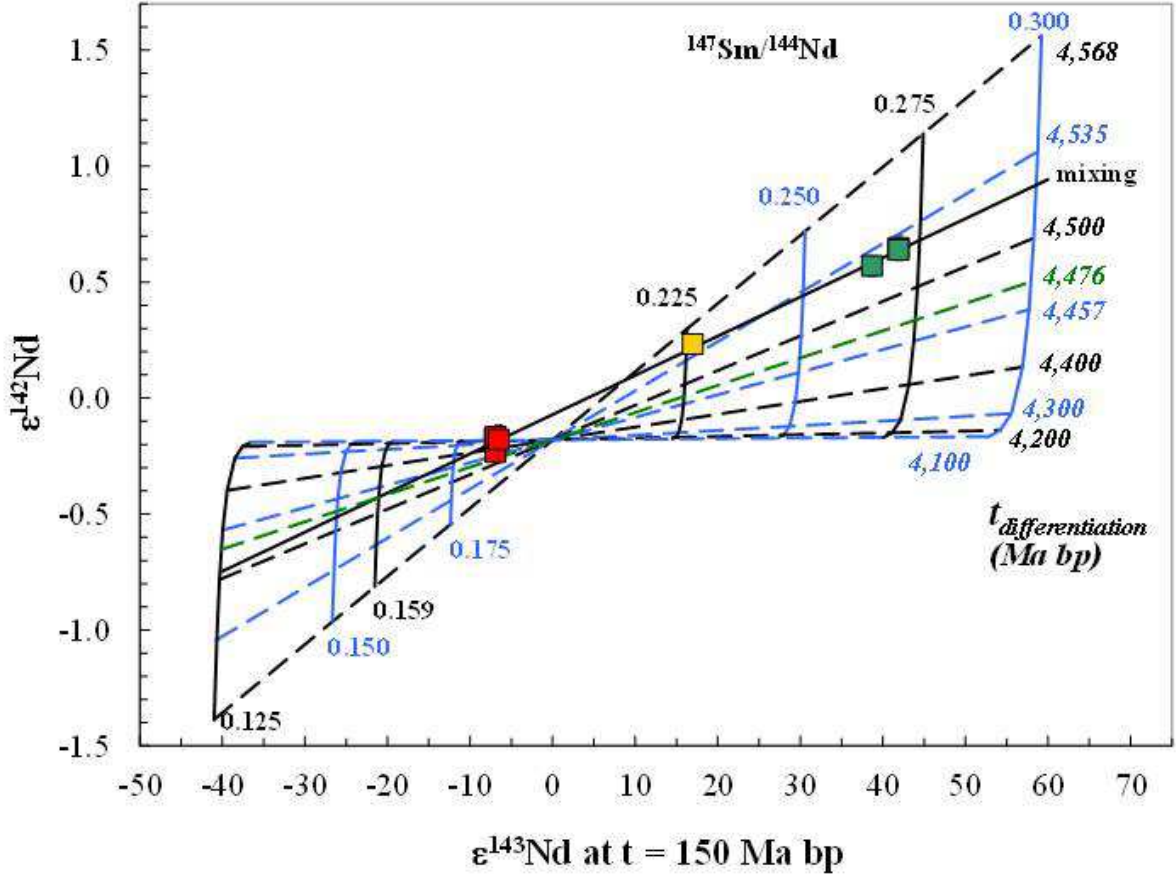

**Fig. S6**

Reinterpretation of the differentiation age of the Martian mantle source of the shergotttite meteorites measured by Debaille et al. (*17*) (solid squares), using the values derived here $t_{1/2}^{146}$ = 68 Ma and $r_0$= ($^{146}$Sm/$^{144}$Sm)$_0$ = 0.0094; compare with their Fig. 2. The axes represent $\varepsilon^{142}$Nd (rel. to the terrestrial standard) vs. $\varepsilon^{143}$Nd

($\varepsilon^{143}Nd(t) = \left( \frac{\left(^{143}Nd/^{144}Nd\right)_t^{DM}}{\left(^{143}Nd/^{144}Nd\right)_t^{CHUR}} - 1 \right) \times 10^4$) at time *t* = 150 Ma before present (bp). Green squares – depleted shergottites, yellow square – intermediate shergottite, red squares – enriched shergottites. Alternating black and blue diagonal dashed lines are loci of equal mantle depletion ages (Ma bp), as labeled. Alternating black and blue solid curves are loci of equal $^{147}$Sm/$^{144}$Nd ratios, as labeled. Lines were calculated with eqs. 1-2. of Debaille *et al.*

(*17*) and their parameters, with the exception of the new values from our work, $t_{1/2}^{146}$ and $r_0 =$ ($^{146}$Sm/$^{144}$Sm)$_0$. The solid black diagonal line is a mixing line between the different shergottite groups. Following Debaille *et al.* (*17*), the differentiation age of the enriched shergottite source is determined by the isochron which passes through the intersection of the mixing line and the $^{147}$Sm/$^{144}$Nd = 0.159 line. In our figure, this is the 4,476 Ma isochron line (dashed green) that corresponds to ~90 Ma after the start of SS formation (taken as 4,568 Ma bp). This analysis assumes that the Sm/Nd ratio of the source is CHUR. Caro *et al.* (*18*) interpreted the mixing line as an isochron and assumed a super-chondritic Sm/Nd ratio for the source. In this case, all shergottite groups came from one depleted Martian mantle source with a differentiation age of ~40 Ma.

**Table S1: Alpha-activity and atom ratios of $^{146}$Sm and $^{147}$Sm in the activated samples.**

| Sample notation | $^{147}$Sm activation | $^{147}$Sm activity (mBq) | $^{146}$Sm activity (mBq) | $^{146}$Sm/$^{147}$Sm activity ratio ($10^{-3}$) | Dilution ratio to AMS sample | $^{146}$Sm/$^{147}$Sm expected atom ratio[a] ($10^{-8}$) | $^{146}$Sm$^{22+}$/$^{147}$Sm$^{22+}$ measured atom ratio ($10^{-8}$) | ($^{146}$Sm/$^{147}$Sm)$_{meas.}$/ ($^{146}$Sm/$^{147}$Sm)$_{expect.}$ |
|---|---|---|---|---|---|---|---|---|
| G-2 | (γ,n) | 46.0 ± 0.1 | 0.185 ± 0.007 | 4.04 ± 0.14 | 137 ± 1 | 2.83 ± 0.16 | 1.90 ± 0.21[b] | 0.670 ± 0.081 |
| P-3 | (p,2nε) | 9.80 ± 0.09 | 2.95 ± 0.05 | 301 ± 6 | 551 ± 8 | 52.7 ± 2.7 | 31.1 ± 2.7[b] | 0.590 ± 0.056 |
| P-4 | | 6.28 ± 0.07 | 2.89 ± 0.05 | 460 ± 10 | 894 ± 10 | 49.6 ± 2.5 | 30.5 ± 2.6[c] | 0.615 ± 0.051 |
| | | | | | | | 31.5 ± 1.3[c] | 0.635 ± 0.027 |
| | | | | | | | 31.4 ± 0.9[c] | 0.634 ± 0.018 |
| | | | | | | | 31.4 ± 0.9[c] | 0.633 ± 0.018 |
| | | | | | | | 29.6 ± 0.6[c] | 0.596 ± 0.011 |
| | | | | | | | 29.2 ± 0.6[c] | 0.589 ± 0.012 |
| | | | | | | | 29.4 ± 0.6[c] | 0.593 ± 0.012 |
| P-4 unw ave ± st dev[d] | | | | | | | 30.5 ± 1.0 | 0.61 ± 0.02 |
| N-2 | (n,2n) | 42.3 ± 0.1 | 0.263 ± 0.006 | 6.27 ± 0.14 | 102 ± 0.6 | 5.92 ± 0.3 | 3.81 ± 0.35[b, c] | 0.643 ± 0.064 |
| | | | | | | | 3.26 ± 0.25[b, c] | 0.551 ± 0.048 |
| | | | | | | | 3.48 ± 0.24[b, c] | 0.588 ± 0.046 |
| N-2 unw ave ± st dev | | | | | | | 3.52 ± 0.28 | 0.59 ± 0.05 |
| N-7 | | | | | 33.2 ± 0.2 | 18.2 ± 0.9 | 12.0 ± 0.3[c] | 0.657 ± 0.031 |
| | | | | | | | 11.3 ± 1.0[c] | 0.622 ± 0.061 |
| | | | | | | | 13.2 ± 0.9[c] | 0.726 ± 0.056 |
| | | | | | | | 11.6 ± 0.2[c] | 0.640 ± 0.027 |
| N-7 unw ave ± st dev | | | | | | | 12.02 ± 0.83 | 0.66 ± 0.05 |
| N-7[e] | | | | | | | 14.8 ± 0.7[c] | 0.81 ± 0.05 |
| | | | | | | | 13.4 ± 1.8[c] | 0.74 ± 0.10 |
| | | | | | | | 14.8 ± 0.6[c] | 0.81 ± 0.05 |
| | | | | | | | 14.6 ± 0.5[c] | 0.80 ± 0.04 |
| N-7[d] unw ave ± st dev | | | | | | | 14.4 ± 0.67 | 0.79 ± 0.05 |
| N-8 | | | | | 92.8 ± 5.0 | 6.51 ± 0.5 | 4.97 ± 0.50[c] | 0.763 ± 0.092 |
| | | | | | | | 4.37 ± 0.42[c] | 0.671 ± 0.078 |
| | | | | | | | 5.03 ± 0.30[b] | 0.772 ± 0.069 |
| | | | | | | | 4.45 ± 0.20[c] | 0.683 ± 0.055 |
| | | | | | | | 3.54 ± 0.39[c] | 0.544 ± 0.070 |
| | | | | | | | 4.08 ± 0.45[c] | 0.627 ± 0.081 |
| | | | | | | | 4.03 ± 0.64[c] | 0.619 ± 0.107 |
| | | | | | | | 4.10 ± 0.39[c] | 0.630 ± 0.073 |
| | | | | | | | 4.87 ± 0.09[c] | 0.748 ± 0.052 |
| | | | | | | | 4.45 ± 0.13[c] | 0.683 ± 0.050 |
| | | | | | | | 4.94 ± 0.21[c] | 0.759 ± 0.060 |

| | | |
|---|---|---|
| N-8 unw ave ± st dev | 5.26 ± 1.18[c] | 0.808 ± 0.189 |
| | 4.51 ± 0.51 | 0.69 ± 0.08 |
| N-8[e] | 3.64 ± 0.13[c] | 0.56 ± 0.04 |
| | 3.66 ± 0.16[c] | 0.56 ± 0.04 |
| | 4.37 ± 0.40[c] | 0.67 ± 0.08 |
| | 4.14 ± 0.65[c] | 0.64 ± 0.11 |
| | 4.94 ± 0.75[c] | 0.76 ± 0.13 |
| | 4.59 ± 0.70[c] | 0.71 ± 0.12 |
| | 4.32 ± 0.15[c] | 0.66 ± 0.05 |
| | 3.75 ± 0.40[c] | 0.58 ± 0.07 |
| | 5.48 ± 0.08[c] | 0.84 ± 0.06 |
| | 4.51 ± 0.75[c] | 0.69 ± 0.12 |
| N-8[e] unw ave ± st dev | 4.34 ± 0.59 | 0.67 ± 0.10 |
| unw ave of ave ±st dev[f] | | 0.66 ± 0.07 |

[a] calculated from cols 3,4, $t_{1/2}(^{146}Sm)$= 103 Ma (*22*,*23*), $t_{1/2}(^{147}Sm)$= 107.0 Ga (*24*)

[b] $^{146}Sm^{22+}/^{152}Sm^{23+}$ measured and converted to $^{146}Sm^{22+}/^{147}Sm^{22+}$ ratio using measured 22+/23+ charge state ratio (Fig. S1)

[c] repeat measurements of the same sample

[d] unweighted average and standard deviation

[e] measured relative to attenuated $^{147}Sm$ count rate (*29*)

[f] unweighted average of sample averages and standard deviation

**Table S2 : Compilation of $^{147}$Sm half-life measurements in the literature**

| Method | Half-life ($10^{11}$ a) | Ref. |
|---|---|---|
| 4π gas flow counter | 1.25 ± 0.06 | Beard et al., 1954 (*39*) |
| Liquid Scintillation | 1.28 ± 0.04 | Beard et al., 1958 (*40*) |
| Ionization Chamber | 1.15 ± 0.05 | MacFarlane et al., 1961 (*41*) |
| Liquid Scintillation | 1.05 ± 0.02 | Wright et al., 1961 (*42*) |
| Liquid Scintillation | 1.04 ± 0.03 | Donhoffer, 1964 (*43*) |
| Ionization chamber | 1.06 ± 0.02 | Gupta et al., 1970 (*44*) |
| Solid state detector | 1.17 ± 0.02 | Kinoshita et al., 2003 (*45*) |
| Liquid scintillation | 1.070 ± 0.009$^a$ | Kossert et al., 2009 (*24*) |
| Track detectors | 1.06 ± 0.01 (metal Sm)<br>1.07 ± 0.01 (Sm$_2$O$_3$) | Su et al., 2010 (*46*) |

$^a$ Value used in this work.